# Auctions: A New Method for Selling Objects with Bimodal Density Functions

Javier Castro[1,2] · Rosa Espínola[1,2] · Inmaculada Gutiérrez[1] · Daniel Gómez[1,2]

## Abstract
In this paper we define a new auction, called the Draw auction. It is based on the implementation of a draw when a minimum price of sale is not reached. We find that a Bayesian Nash equilibrium is reached in the Draw auction when each player bids his true personal valuation of the object. Furthermore, we show that the expected profit for the seller in the Draw auction is greater than in second-price auctions, with or without minimum price of sale. We make this affirmation for objects whose valuation can be modeled as a bimodal density function in which the first mode is much greater than the second one. Regarding the Myerson auction, we show that the expected profit for the seller in the Draw auction is nearly as good as the expected profit in the optimal auction, with the difference that our method is much more simple to implement than Myerson's one. All these results are shown by computational tests, for whose development we have defined an algorithm to calculate Myerson auction.

## 1 Introduction

The importance and widespread use of auctions is well established. A high volume of economic transactions is conducted through different types of auctions in both the private and public spheres. Auctions have been used in the selling of goods and services for thousands of years (Kagel and Roth 1995; Krishna 2010; Lorentziadis 2016; Milgrom and Weber 1982; Zhou and Li 2019).

---

The classical auction model, known as benchmark model, is based on four principal assumptions: the buyers are risk neutral (An et al. 2011; Holt 1980); buyers' values are independent and private (Winkler and Brooks 1980; Yamaji et al. 2016); the buyers are symmetric (Gayle and Richard 2008); and the payment is a function of the bids alone (see Boyer and Brorsen 2014; Wolfstetter 1996 for more details and other additional technical assumptions). The benchmark model is adopted in this paper for theoretical results. Hence, we assume that, in the absence of any knowledge of the opposite, all bidders will make offers maximizing their own profit (García 2013; Vickrey 1961). Several examples of auctions which are not based on benchmark model have been analyzed in (Dutting et al. 2017; Fibich and Gavish 2012; Hubbard and Kirkegaard 2013; Ji and Li 2016; Peng and Yang 2010; Xu et al. 2019).

There are several auction methods (Budde and Minner 2014; Chen et al. 2014; Erlanson 2014; Hailu and Thoyer 2010; Milgrom and Weber 1982). Two of the most popular auctions are the second-price auction and the second price with minimum price of sale auction, but this popularity does not guarantee an optimal profit for the seller. In fact, it is generally realized that, taking into account the standard assumptions, and from the seller's point of view, Myerson auction (Myerson 1981) is the optimal auction method.

We focus on situations related to the sale of objects whose valuation can be modeled with a bimodal density function. Myerson auction, which has a minimum price of sale, usually defines a draw interval in these cases. This auction involves a complex process of understanding. It has not been studied in depth over the literature (Dudik et al. 2017; Morgenstern and Roughgarden 2015; Pierrakos 2013). Nevertheless, we can find some works related to different approaches of the Myerson auctions in certain specific scenarios, for example in cases with resale (Zheng 2002) or on a Bayesian context (Deng and Zhu 2018) among many others (Lee et al. 2021a, b). In this article, it has been crucial to obtain the Myerson auction explicitly for this situation.

On the other hand, we introduce the so-called Draw auction. This auction has two parameters that are selected before starting the sale of the object: a fixed price $c$, and an integer $k \in \{1,\ldots,n-1\}$, where $n$ is the number of bidders in the auction. In this auction the object is assigned to one buyer as follows. If more than one buyer is willing to pay this fixed price, $c$, then the object is assigned to the buyer with the highest bid at the price of the second highest bid. If only one buyer is willing to pay $c$, then the object is assigned to this buyer, who has to pay this fixed price, $c$. If none of the buyers is willing to pay the fixed price $c$, then the object is assigned randomly (by a draw) to one of the $k$ buyers with the highest bids. So, the winner of the draw gets the object, paying for it the price of the $(k+1)$-th highest bid, $x_{(k+1)}$, where $k$ is less than the number of buyers in the auction, $n$.

The remainder of the paper is organized as follows. In Sect. 2, we describe the background of the paper. We give the definition of some important concepts related to auctions. We also describe several realistic situations in which it would be useful the application of the new auction. We define a polynomial algorithm to compute the Myerson auction for the case of bimodal distributions (with uniform distributions) in Sect. 3. We introduce the Draw auction in Sect. 4. In Sect. 5, we present some computational results. We carry on with some conclusions in Sect. 6. All the details about the notation, the full demonstration of the propositions and the pseudocode of the algorithms are provided in the "Appendices A, B and C".

## 2 Background Description

### 2.1 Auctions

In this subsection we provide the definition of some basic concepts about auctions, which will be needed for following up and understanding of the paper.

First, let us explain some important concepts that are needed to determinate the optimal strategy in an auction. These definitions and many others related to them, can be found with more details in Gibbons (1992).

- **Dominant Strategy**: let $G$ be the static Bayesian game $\mathcal{G} = \{S_1, \ldots, S_n; F_1, \ldots, F_n; p_1, \ldots, p_n; u_1, \ldots, u_n\}$, where $S_i$ is player $i$'s action or strategy space; $u_i(s_1, \ldots, s_n; w_i)$ is player $i$'s payoff when it chooses the actions $(s_1, \ldots, s_n)$, and $w_i$ is player $i$'s type, which is privately known by himself, and belongs to a set of possible types, $W_i$. The vector $(p_1, \ldots, p_n)$ represents players' beliefs, when the belief of player $i$, $p_i(w_{-i} \mid w_i)$, describes $i$'s uncertainty about the $n-1$ other players' possible types, $w_{-i}$, given his own type, $w_i$. A strategy for player $i$ is a function $s_i(w_i)$ where, for each type $w_i \in W_i$, $s_i(w_i)$ specifies the action from the feasible set $S_i$ that type $w_i$ would choose if drawn by nature. We define the expected value as $E_i[s_1, \ldots, s_n] = \sum_{w_{-i} \in W_{-i}} u_i(s_1, \ldots, s_n; w) p_i(w_{-i} \mid w_i)$.
  Strategy $s'_i$ is dominated by strategy $s''_i$ if $\forall w_i \in W_i$, for each feasible combination of the other players' strategies, $i$'s payoff from playing $s'_i$ is less than $i$'s payoff from playing $s''_i$:
  
  $$E_i[s_1 \ldots, s_{i-1}, s'_i, s_{i+1} \ldots, s_n] \leq E_i[s_1 \ldots, s_{i-1}, s''_i, s_{i+1} \ldots, s_n]$$
  
  for each $(s_1 \ldots, s_{i-1}, s_{i+1} \ldots, s_n)$ that can be constructed from the other players' strategy spaces $S_1, \ldots, S_{i-1}, S_{i+1}, \ldots, S_n$.

- **Bayesian Nash Equilibrium (BNE)**: let $\mathcal{G}$ be the static Bayesian game $\mathcal{G} = \{S_1, \ldots, S_n; W_1, \ldots, W_n; p_1, \ldots, p_n; u_1, \ldots, u_n\}$. The strategies $s^* = (s^*_1, \ldots, s^*_n)$ are a (pure-strategy) *Bayesian Nash Equilibrium* if for each player $i$ and for each of $i$'s types, $w_i \in W_i$, $s^*_i(w_i)$, solves:
  
  $$\max_{s_i \in S_i} E_i[s^*_1, \ldots, s^*_{i-1}, s_i, s^*_{i+1}, \ldots, s^*_n].$$

That is, no player wants to change his strategy, even if the change involves only one action by one type.

In addition, it is important to take into account that a Dominant Strategy implies a Bayesian Nash Equilibrium (Carbonell-Nicolau and McLean 2018; Einy et al. 2002).

On the other hand, these are three of the most popular auctions will be taken into account in following pages (in addition to the new one defined, the **Draw auction**): **second-price auction**, **second-price with minimum price of sale auction** and **Myerson auction**.

The first two auctions are well known (Mares and Swinkels 2014). In both, the best bidder gets the object, and he has to pay the second highest offer. The only difference between both auctions is that the second method establishes a minimum price of sale, and the object is not sold for a price lower than it.

On the other hand the optimal auction model was proposed by R. Myerson in Myerson (1981). The reader can confirm the complexity associated with the understanding of this auction. In auction literature, it is normally assumed that the bid price is a continuous random variable. For defining his model, Myerson denotes as $F$ the distribution function and as $f$ the density function of this random variable. Then, he defines some auxiliary functions: $h(q) = F^{-1}(q) - \frac{1-q}{f(F^{-1}(q))} = c(F^{-1}(q))$ and $H(q) = \int_0^q h(r)dr$.

Once these equations are defined, Myerson calculates the convex hull of the function $H$, called $G$, and its derivative, $g(q) = G'(q)$.

Finally, if player $i$ offers the bid $x_i$, Myerson defines: $\overline{c}(x_i) := g(F(x_i))$, and $M(x) := \{i \mid 0 \leq \overline{c}(x_i) = max_{j \in N}(\overline{c}(x_j))\}$.

Next, Myerson affirms that $(\overline{p}, \overline{x})$ is an optimal auction, in which the probability that the object is assigned to a bidder is defined by the vector $\overline{p}(x) = (\overline{p}_1(x), \ldots, \overline{p}_n(x))$, where $\overline{p}_i(x) = \frac{1}{|M(x)|}$, if $i \in M(x)$, and 0 otherwise; and the function that defines the expected payment of the object is $\overline{x}(x) = (\overline{x}_1(x), \ldots, \overline{x}_n(x))$, where $\overline{x}_i(x) = \overline{p}_i(x)x_i - \int_{a_i}^{x_i} \overline{p}_i(x_{-i}, s_i)ds_i$.

Let us note that in Myerson auction and in second-price with and without minimum price of sale auctions, the BNE is achieved when each player bids his own valuation of the object, $v_i$.

## 2.2 Objects with Bimodal Density Functions

In this subsection, we describe several types of objects whose valuation can be modeled as a random variable with a bimodal density function in which the first mode is much greater than the second one (see Fig. 1).

The valuation of each object is subjective and may therefore have a 'low' value (first mode) for most buyers, and a 'high' value (second mode) for some of the buyers. Four groups of objects that exhibit bimodal valuations are:

1. Objects with possible personal value: for buyers with a personal connection to the object, it will have a 'high' value. For the remainder of the buyers, the object will have a 'low' value. Some examples include a home, jewels or more generally, any objects that have been seized or inherited. A family member or friend may be particularly interested in these objects.
2. Objects whose appraisal is difficult: most people will act with caution so they will assign a 'low' value to the object. However, some buyers may view the object as a worthwhile business investment and, therefore, they will assign a

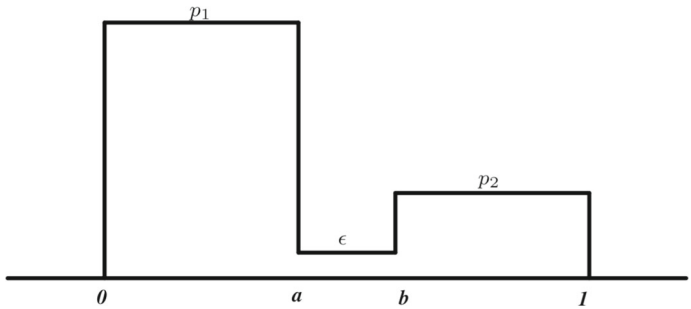

**Fig. 1** Example of a bimodal distribution defined by a convolution of uniforms

'high' value to the object. Some examples of this type of object are a telecommunications license or seized company.

3. Artistic objects: anyone who does not 'believe' in the artistic merit of the object will assign a 'low' valuation to it. However those who judge the object as artistically superior will assign a 'high' valuation to the object. Some examples of artistic object include works of art, antiques and collectible objects such as stamps and coins.
4. Objects about which some buyers have privileged information: if a rumor exists concerning an object (this concrete rumor will increase the value of the object if it is true), then those without confirmation of the rumor will assign a 'low' valuation to the object. However, buyers with confirmation of the rumor (privileged information) will assign a 'high' valuation. Some examples of this type of objects are the purchase of a solar energy company when there is a rumor of future government aid to the sector, or the purchase of a home when there is a rumor concerning the construction of a nearby subway station.

These situations satisfy the assumption of symmetric buyers as, neither the seller nor the remainder of the buyers know a priori whether a buyer has a 'high' or 'low' valuation of the object. Suppose that each of the $n$ buyers has a probability $(1-p)$ of having a 'low' valuation, and a probability $p$ of having a 'high' valuation of the object. Then, we can construct $n+1$ scenarios in which there are $k$ 'high' buyers' valuation and $n-k$ 'low' buyers' valuation (for $k$ between 0 and $n$), with probabilities $\frac{n!}{(n-k)!k!}(1-p)^{(n-k)}p^k$. Another situation, different than the previous one, occurs when one or more 'high' buyers' valuation are known to be in the auction (Kagel and Levin 1999; Kirkegaard 2012).

## 3 Computing the Myerson Auction in Bimodal Models

This section has three different parts: first of all, we propose a method to calculate Myerson auction for the case of a bimodal distribution defined by a convolution of uniforms in which the first mode is much greater than the second one. We summarize this method in the Algorithm *BU Myerson*. Then, we define a proposition to affirm that the method *BU Myerson* provides the result of Myerson auction for described situation. Finally, we give an sketch of the proof of this proposition, whose detailed explanation can be found in the "Appendix B".

Myerson auction involves a complex process of understanding, which has not been deeply studied over the literature (Dudik et al. 2017; Morgenstern and Roughgarden 2015; Pierrakos 2013). To carry on with this article, it has been crucial to obtain the Myerson auction explicitly for the case of a bimodal distribution defined by a convolution of uniforms with $0 < a < b < 1$ (see Fig. 1 and function (1)). We work under the assumptions of benchmark model.

$$f(q) = \begin{cases} \dfrac{p_1}{a} & if \quad q \in (0, a] \\ \dfrac{\epsilon}{(b-a)} & if \quad q \in (a, b] \\ \dfrac{p_2}{(1-b)} & if \quad q \in (b, 1) \end{cases} \quad (1)$$

Myerson auction usually behaves like a second-price auction, but in some circumstances, it defines a draw interval, used to designate the player who takes the good and the price he has to pay for it (and the price the rest of the buyers would have to pay). Due to this behavior, the calculation of the auction becomes a hard process.

In this section we define a method to calculate in an efficient way (complexity $O(n)$) the auction proposed by Myerson (1981) for mentioned situation. The input variables of this procedure are: $a$ (top limit of the first mode), $b$ (lower limit of the second mode), $\epsilon$ (cumulative probability between both modes), $p_1$ (cumulative probability of the first mode), $p_2$ (cumulative probability of the second mode), $x$ (bid vector). The output variables are $\bar{p}$ and $\bar{x}$, both described in Sect. 2.1. Let us describe the method *BU Myerson* step by step. Its pseudocode can be found in the "Appendix C", Algorithm 1. After the definition, we assert that the application of this method provides the solution of the Myerson auction when the density function is similar to function (1).

**Method BU Myerson**

The starting points are the vector $(x_{(1)}, \ldots, x_{(n)})$, which is the bidding vector $x = (x_1, \ldots, x_n)$ with descending order, and the density function $f(q)$ previously introduced. The final objective is to obtain the vectors $\bar{p} = (\bar{p}_{(1)}, \ldots, \bar{p}_{(n)})$ and $\bar{x} = (\bar{x}_{(1)}, \ldots, \bar{x}_{(n)})$, which define the probability that the object is assigned to a particular buyer $i$, and the price every buyer has to pay, respectively. Unless stated otherwise, the value of all the components of these vectors will be initially zero. Let us remark that, because of our initial assumption about these vectors, on the following we will just specify those components in which there is a modification. Then, to calculate the final value of these two vectors by means of the Myerson auction, we have to obtain these four parameters:

- **$x_{min}$**. First minimum price of sale.
- **$x_{\ell\ell}$**. Lower limit of the Myerson draw.

- $x_{cut}$. Upper limit of the Myerson draw or second minimum price of sale.
- $\beta_0$. Bound needed to fix the minimum sale value, . If $\beta_0 < 0$, the minimum selling price is $x_{cut}$. Otherwise, this minimum price is $x_{min}$.

Once these values have been calculated, there are several possible scenarios to determine the allocation of the Myerson auction. These cases are the following:

- When $\beta_0 < 0$, the minimum price sale is $x_{cut}$.
  - If $x_{(1)} \geq x_{cut}$ and $x_{(2)} \in [0, x_{cut})$, then $\overline{p}_{(1)}(x) = 1$ and $\overline{x}_{(1)}(x) = x_{cut}$. In this situation there is only one buyer who is willing to pay the minimum price $x_{cut}$. Then, this price $x_{cut}$ is fixed as the selling price.
  - Else, if $x_{(2)} \geq x_{cut}$, then $\overline{p}_{(1)}(x) = 1$ and $\overline{x}_{(1)}(x) = x_{(2)}$. In this situations there are at least two buyers who are willing to pay the minimum price of sale, $x_{cut}$. Then, the selling price is the second highest bid, $x_{(2)}$.
  - Else, $x_{(1)} < x_{cut}$, so $\overline{p} = (0, \ldots, 0)$ and $\overline{x} = (0, \ldots, 0)$. In this situation, none of the buyers are willing to pay the minimum price of sale $x_{cut}$, so the object is not sold.

- When $\beta_0 \geq 0$, the minimum price sale is $x_{min}$.
  - If $x_{(1)} \notin [0, x_{min})$ and $x_{(2)} \in [0, x_{min})$, then $\overline{p}_{(1)}(x) = 1$ and $\overline{x}_{(1)}(x) = x_{min}$. In this situation there is only one buyer who is willing to pay the minimum price $x_{min}$. Then, this price $x_{min}$ is fixed as the selling price.
  - Else, if $x_{(1)}, x_{(2)} \in [x_{min}, x_{\ell\ell})$, then $\overline{p}_{(1)}(x) = 1$ and $\overline{x}_{(1)}(x) = x_{(2)}$. In this situation there are at least two buyers who are willing to pay the minimum price of sale, $x_{min}$, but none of them offer a bid higher than the lower limit of the draw, $x_{\ell\ell}$. Then, the selling price is the second highest bid, $x_{(2)}$.
  - Else, if $x_{(1)} \in (x_{\ell\ell}, x_{cut}]$ and $x_{(2)} \in [x_{min}, x_{\ell\ell})$, then $\overline{p}_{(1)}(x) = 1$ and $\overline{x}_{(1)}(x) = x_{(2)}$. In this situation there are at least two buyers who are willing to pay the minimum price of sale, $x_{min}$, but only one of them offers a bid higher than the minimum limit of the draw, $x_{\ell\ell}$. Then, the selling price is the second highest bid, $x_{(2)}$.
  - Else, if $x_{(1)}, x_{(2)} \in (x_{\ell\ell}, x_{cut}]$, then $\overline{p}_{(i)}(x) = \frac{1}{|\mathcal{D}|}$ and $\overline{x}_{(i)}(x) = \frac{x_{\ell\ell}}{|\mathcal{D}|}$, $\forall i \in \mathcal{D}$, being $\mathcal{D}$ the set of buyers whose bid belongs to the draw interval. In this situation there are at least two buyers whose bids are in the draw interval, (and not higher than its upper limit $x_{cut}$), so the object will be raffled among all the buyers belonging to $\mathcal{D}$. In this case, the selling price is $x_{\ell\ell}$, the lower limit of the draw interval.
  - Else, if $x_{(1)} > x_{cut}$ and $x_{(2)} \in (x_{\ell\ell}, x_{cut}]$, then $\overline{p}_{(1)}(x) = 1$ and $\overline{x}_{(1)}(x) = x_{cut} - \frac{x_{cut} - x_{\ell\ell}}{|\mathcal{D}|+1}$, being $\mathcal{D} = \{i \mid x_i \in [x_{\ell\ell}, x_{cut}]\}$. In this situation there is one bid higher than the upper limit of the draw interval, $x_{(1)} > x_{cut}$. There are also some buyers $j$ who offer a bid belonging to the draw interval, so $j \in \mathcal{D}$. In order to avoid the raffle, the highest bidder will have to pay the upper limit of the draw interval without the expected benefit in case it was in $\mathcal{D}$, i.e. $x_{(1)}(x) = x_{cut} - \frac{x_{cut} - x_{\ell\ell}}{|\mathcal{D}|+1}$.

- Else, if $x_{(1)} > x_{cut}$, then $\overline{p}_{(1)}(x) = 1$ and $\overline{x}_{(1)}(x) = x_{(2)}$. In this situation there is one bid higher than the upper limit of the draw interval, $x_{(1)} > x_{cut}$. The rest of the bids are outside of the draw interval, so the selling price is the second highest bid, $x_{(2)}$.
- Else, $x_{(1)} < x_{min}$, so $\overline{p} = (0,\ldots,0)$ and $\overline{x} = (0,\ldots,0)$. In this situation in which none of the buyers are willing to pay the minimum price $x_{min}$ the object is not sold.

An sketch of the process to calculate those four parameters, $x_{min}$, $x_{\ell\ell}$, $x_{cut}$, $\beta_0$ is showed below. The justification for this process can be found in the extended proof of the Proposition 1, showed in the "Appendix B", Proof 1. The implementation of this process is showed in the Algorithm 1, included in the "Appendix C".

- $F(q)$ is the distribution function of $f(q)$.
- $A = \frac{p_1}{p_2}\frac{1-b}{a}$
- $B = -[\frac{p_1}{2a}(b - \frac{(1-b)}{p_2}(1 + p_1 + \epsilon)) + \frac{1}{2}]$
- $C = \frac{-a}{p_1}A^2 + \frac{1-b}{p_2}$
- $D = 2\frac{a}{p_1}AB$
- $E = \frac{-a}{p_1}B^2 + b - \frac{1-b}{p_2}(p_1 + \epsilon)$
- $y_1 = \frac{-D+\sqrt{D^2-4CE}}{2C}$
- $y_2 = \frac{-D-\sqrt{D^2-4CE}}{2C}$
- $z_1 = Ay_1 - B$
- $z_2 = Ay_2 - B$
- $z_3 = (p_1 + \epsilon) - \sqrt{(p_1 + \epsilon)^2 - (p_1 + \epsilon) - \frac{p_1 b}{a}(p_1 + \epsilon - 1)}$
- $H_1(q) = \frac{a}{p_1}(q^2 - q)$
- $H_3(q) = b(q - 1) + \frac{1-b}{p_2}(q^2 - (1 + p_1 + \epsilon)q + (p_1 + \epsilon))$
- If exists $z_1 \in [0, p_1]$ andexists $y_1 \in [p_1 + \epsilon, 1] \implies z_0 := z_1$ and $y_0 := y_1$.
- Otherwise, if exists $z_2 \in [0, p_1]$ andexists $y_2 \in [p_1 + \epsilon, 1] \implies z_0 := z_2$ and $y_0 := y_2$.
- Otherwise, if exists $z_3 \in [0, p_1] \implies z_0 := z_3 y_0 := (p_1 + \epsilon)$.
- Otherwise $\implies z_0 := 0$ and $y_0 := (p_1 + \epsilon)$.

Then,

- $x_{cut} = F^{-1}(y_0)$
- $\beta_0 = \frac{H_3(y_0) - H_1(z_0)}{y_0 - z_0}$
- $x_{min} = \frac{a}{2p_1}$
- $x_{\ell\ell} = \frac{az_0}{p_1}$

**Proposition 1** *The method previously described, summarized in the Algorithm BU Myerson, provides the result of Myerson auction for the case of a Bimodal distribution defined by a convolution of Uniforms described in the Eq. (1) in which the first mode is much greater than the second one.*

***Proof*** The proof of this proposition is deeply explained in the "Appendix B", Proof 1. The first step consists on the calculation of the function $H(q) = \int_0^q h(r)dr$, where $h(q) = F^{-1}(q) - \frac{1-q}{f(F^{-1}(q))}$, $f$ is the density function and $F$ is the distribution function. $H$ is a piecewise function with three parts, $H_1$, $H_2$ and $H_3$, whose convex hull is called $G$. The characterization of $G$ depends on the calculation of several lines:

1. The tangent lines among $H_1$ and $H_3$.
2. The tangent line to $H_1$ which passes through the point $((p_1 + \epsilon), H_3(p_1 + \epsilon))$.
3. The line that passes through the points (0, 0) and $((p_1 + \epsilon), H_3(p_1 + \epsilon))$.

To carry on with these points, several auxiliary steps must be completed, all of them based on the calculation of some lines, for which we know either a couple of points through which each one passes, or a direction and a point of it.

Once these three points are fixed, we have to work in the calculation of the points $z_0$, $y_0$, $\alpha_0$ and $\beta_0$, which will fix the definition of the convex hull of $H$, $G$. Finally, the characterization of the optimum bid defined by Myerson, depends on the definition of the function $G$. □

## 4 The Draw Auction

For each buyer $i$, with $i \in \{1, \ldots, n\}$, we denote the $i$-th valuation of the object by $v_i$ and the personal valuation of the seller by $v_0$. We represent the bids of the buyers by the vector $(x_1, \ldots, x_n)$, and we denote the ordered list of bids as $(x_{(1)}, \ldots, x_{(n)})$, where $x_{(1)}$ is the highest bid, $x_{(2)}$ is the second highest bid and so on. Similarly, $(v_1, \ldots, v_n)$ denotes the vector of buyer valuations, and $(v_{(1)}, \ldots, v_{(n)})$ denotes the ordered vector of valuations, where $v_{(1)}$ is the highest valuation, $v_{(2)}$ is the second highest valuation and so on.

Using this notation, then we define a new auction with good behavior when the valuations can be modeled as a random variable with a bimodal density function in which the first mode is much greater than the second one, and satisfying the assumptions of the benchmark model. Hereafter, the new auction is called **Draw auction**, and its formal definition is described below, where we propose a step-by-step method to calculate this new auction, whose input variables are $k$, $c$, $x$, and output variables are $\bar{p}$ and $\bar{x}$. The corresponding pseudocode can be found in the "Appendix C", Algorithm 2.

1. We select a value $k \in \{1, \ldots, n-1\}$, and a constant value $c > v_0$, which represents a fixed price.
2. Knowing the values of $k$ and $c$, each buyer places a bid, $x_i$.
3. To assign the object to a single buyer, we have to distinguish among three cases:

(a) If more than one buyer is willing to pay the price $c$, that is, if the second highest bid, $x_{(2)}$, satisfies the inequality $x_{(2)} - c > \frac{1}{k}\left(x_{(2)} - x_{(k+1)}\right)$ (and therefore $x_{(1)}$ also satisfies this inequality due to $x_{(1)} \geq x_{(2)}$), then the object is assigned to the buyer who has proposed the highest bid, $x_{(1)}$, and this buyer pays $x_{(2)}$ units for the object.

(b) If only one buyer is willing to pay the price $c$, that is, if the second highest bid, $x_{(2)}$, does not satisfy the inequality $x_{(2)} - c > \frac{1}{k}\left(x_{(2)} - x_{(k+1)}\right)$, but the highest bid, $x_{(1)}$, does satisfy the inequality $x_{(1)} - c > \frac{1}{k}\left(x_{(1)} - x_{(k+1)}\right)$, then the object is assigned to the buyer who has proposed the highest bid, $x_{(1)}$, and this buyers pays $c$ units for the object.

(c) If none of the buyers is willing to pay the price $c$, that is, if the highest bid does not satisfy the inequality $x_{(1)} - c > \frac{1}{k}\left(x_{(1)} - x_{(k+1)}\right)$, then the object is assigned randomly (by a draw) to one of the buyers with the $k$ highest bids. The winner of the draw pays $x_{(k+1)}$ units for the object.

Note that the second-price auction is a particular case of the draw auction. When $k = 1$, the results of both auctions are the same for any bid. The following example illustrate the behavior of the Draw auction.

**Example 1** Let us consider a Draw auction with $n = 4$, $k = 2$ and $c = 0.4$.

- If the bid vector is (0.7, 0.6, 0.3, 0.1), then more than one buyer is willing to pay the price $c$ (implying that the first and second buyers satisfy the inequality). In this case the object is assign to the first buyer at a cost of 0.6.
- If the bid vector is (0.7, 0.45, 0.3, 0.1) or (0.7, 0.3, 0.3, 0.1), then only one buyer is willing to pay the price $c$ (implying that the first bid satisfies the inequality, but the second one does not satisfy it). In this case the object is assign to the first buyer at a cost of 0.4.
- If the bid vector is (0.45, 0.4, 0.3, 0.1), then none of the buyers is willing to pay the price $c$ (implying that no bid satisfies the inequality). In this case, the object is assigned randomly by a draw between the first and the second buyer. The amount paid by the winning buyer in the draw is 0.3.

When working with auctions, the main objective is to achieve the BNE. We next prove that the BNE strategy in the Draw auction coincides with the valuation vector. This result shows the efficiency and simplicity of the new defined auction.

**Proposition 2** *In the Draw auction, the bid vector composed by buyers' valuations is a dominant strategy and therefore it is a BNE strategy.*

**Proof** The demonstration of this proposition is explained in detail in the "Appendix B", Proof B. Our starting point is the fact the every dominant strategy is also a BNE strategy. Hence, we demonstrate that the bid vector defined by buyers' valuations is a dominant strategy. To carry on with it, we use the auxiliary vector $x^* = (x_1, \ldots x_{i-1}, v_i, x_{i+1}, \ldots, x_n)$, and we demonstrate that, $\forall x \in \mathfrak{R}_+^n$, $E_i[x^*] \geq E_i[x]$. This comparison between these expected values is performed by considering several feasible cases, based on eight Yes/Not questions. This analysis has to be done both for the case in which $x_i \geq v_i$, and for the case in which $x_i < v_i$. □

An alternative and equivalent definition of the Draw auction is as follows:

1. We specify a value $q \in [0, 1]$, and a constant $c > v_0$. We define $k = [qn]$. This value will have the same interpretation than it has in the original auction.
2. The auction begins with a very low starting bid, and this bid is gradually increased. Buyers who are not prepared to pay the new bid must leave the auction.
3. When $(n - k)$ buyers have left the auction, leaving $k$ buyers, each of the remaining buyers is asked (privately) weather he is willing to pay a price of $c$ for the object. We must distinguish three cases:

   (a) If more than one buyer is agree to pay $c$, then the seller continues the auction from a starting price of $c$ with the set of buyers who agree to pay $c$. The buyer who leaves last receives the object. The price of the object for this buyer is the bid at the moment when the penultimate buyer leaves, $x_{(2)}$.
   (b) If only one buyer agrees to pay $c$, then this buyer receives the object at a price of $c$.
   (c) If none of the buyers agrees to pay $c$, then a draw is performed among the $k$ buyers. In this case, the price of the object is the bid at the moment when $(n - k)$ buyers have left, $x_{(k+1)}$.

This new definition is equivalent to the original one, as both auctions always yield the same result if the buyers employ their BNE strategies. In the alternative definition, the BNE strategy is to remain in the auction until the bid is equal to $v_i$ and the price $c$ is accepted if and only if $v_i - c > \frac{1}{k}(v_i - x_{(k+1)})$.

We conclude this section with two technical remarks:

- If we assume that $n$ is unknown at the moment in which $k$ is designated, then a function $k(n)$ should be defined in place of the single integer $k$, such that an integer between 1 and $n - 1$ is assigned for each possible value of $n$.
- It is straightforward to generalize the Draw auction by defining a function $c(x_{(k+1)}, \ldots, x_{(n)})$, which takes the lowest bids as arguments, rather than selecting a constant $c$. This generalization may be useful when the valuations (and therefore the bids) of the buyers are correlated.

These changes in the definitions of $k$ and $c$ do not alter the conclusion that the BNE strategy is the bid vector composed by buyers' valuations of the object.

## 5 Computational Results

This section has three different parts. First, we explain the background of the computational tests, with a detailed explanation of the application scenarios and the parameters. Then, we test the efficiency of the Draw auction by comparing it with the optimal one (Myerson auction). Finally, we show that, from seller viewpoint, the Draw auction is better than those methods which are usually applied in real life situations, second-price auctions, with and without minimum price of sale, both in original scenarios and robustness analysis.

### 5.1 The Computational Parameters

Some computational tests have been performed for various objects with subjective valuations. These valuations can be modeled as random variables with bimodal density functions assuming the following forms:

- A density function $f_1(x)$ on the interval $[a_1, a_2]$, with a high probability $(1 - p)$ of assigning a low valuation.
- A density function $f_2(x)$ on the interval $[b_1, b_2]$, with a low probability $p$ of assigning a high valuation.

The functions $f_1(x)$ and $f_2(x)$ could be any density function, for example, the beta distribution, the uniform distribution or the normal distribution. Particularly in this section we assume that the seller's personal valuation is zero, $v_0 = 0$, and we consider the density function defined in Fig. 1, where $p_1 = (1 - p)$; $\epsilon = 0$; $p_2 = p$; $a_1 = 0$; $a_2 = a$; $b_1 = b$, and $b_2 = 1$. The behavior is expected to be similar for any density function similar in appearance, i.e., any density function with two clearly defined modes.

The application of the Draw auction has a special interest in the auction of objects whose valuation can be modeled as a bimodal density function in which the first mode is much greater than the second one. Specifically, the Draw auction works well in situations in which a player appearing in the second mode is not the most likely situation, but it is not disposable. This is the general situation in the sale of objects such as those described in Sect. 2.2. To quantify these scenarios, we have only considered the cases in which the expected value of a bid in the second mode, $(np)$, is equal or less than 0.5.

The values used in our computations are $np = 0.1, 0.2, 0.3, 0.4$ and $0.5$; $\epsilon = 0$; $a = 0.2$ and $0.4$; and $b = 0.6$ and $0.8$, as it shown in Tables 2 and 3. The values of $k$ and $c$ in the Draw auction and $v^*$ in the second-price with minimum price of sale auction, will be determined by maximizing the expected profit for the seller.

Given $k$ and $c$, the expected profit for the seller is calculated as follows for the Draw auction, assuming that the buyers employ the BNE strategy:

$$E[Y] = P(Z=0)E[v_{(k+1)}/Z=0] + P(Z=1)c + \sum_{i=2}^{n} P(Z=i)E[v_{(2)}/Z=i], \quad (2)$$

where $Z$ is a random variable which represents the number of valuations that satisfy the inequality $v_i - c > \frac{1}{k}(v_i - v_{(k+1)})$.

On the other hand, given $v^*$, if the buyers use the BNE strategy, the expected profit for the seller in the second-price with minimum price of sale auction is calculated as follows:

$$E[Y] = P(W=1)v^* + \sum_{i=2}^{n} P(W=i)E[v_{(2)}/W=i], \quad (3)$$

where $W$ is a random variable which represents the number of valuations that satisfy the inequality $v_i > v^*$.

The calculation of the auction parameters ($v^*$ or $c$ and $k$), that yield the highest expected profit for the seller in the second-price with minimum price of sale auction and Draw auction depends on the density function used to model the valuation of the object. This calculation is usually difficult, therefore, it is necessary to use heuristic methods. A basic heuristic method has been employed in this paper. For the draw auction, we constructed a grid with all of the possible values of $k$ (i.e. $k = 1, \ldots, n-1$) and values of $c$ between 0 and 1 with an increment of 0.01 (i.e. $c = 0, 0.01, \ldots, 0.99, 1$). For the second-price with minimum price of sale auction, the grid included values of $v^*$ between 0 and $a$, and between $b$ and 1, with an increment of 0.001 (i.e. $v^* = 0, 0.001, \ldots, a - 0.001, a, b, b + 0.001, \ldots, 0.999, 1$). The others two auctions shown in computational results do not have parameters in theirs definition. In both cases, the expected profit for the seller was calculated on the grid using 1000000 simulations.

In order to set light to the methodological process to estimate the parameters, we explain the method used by means of its pseudocode, in the Algorithm 3 provided in the "Appendix C".

To define an auction, it is important to take into account that, if it is applied in real situations, in general, there is a period of time in which the auction application scenario changes with regard to the moment when it was defined. For this reason, we study the behavior of the Draw auction on two the most used auctions (second-price auctions, with or without minimum price of sale) in 21 different scenarios: the first scenario is the initial case, based on the assumptions of the benchmark model, without variations in different parameters of the auction or the situation in which it is applied; the others 20 different scenarios are related to a robustness analysis, and they have variations in intervals of the first mode.

The analysis of robustness has been focused in the first mode, as it is logical to think that the second mode, which represents 'subjective' valuations, will be less affected along the time. So, if there is a person who is convinced to offer a high bid due to a personal reason, this bid will not be modified even if, objectively and for the rest of player, the valuation of the object to be sold changes from the moment when the auction is defined until it is done.

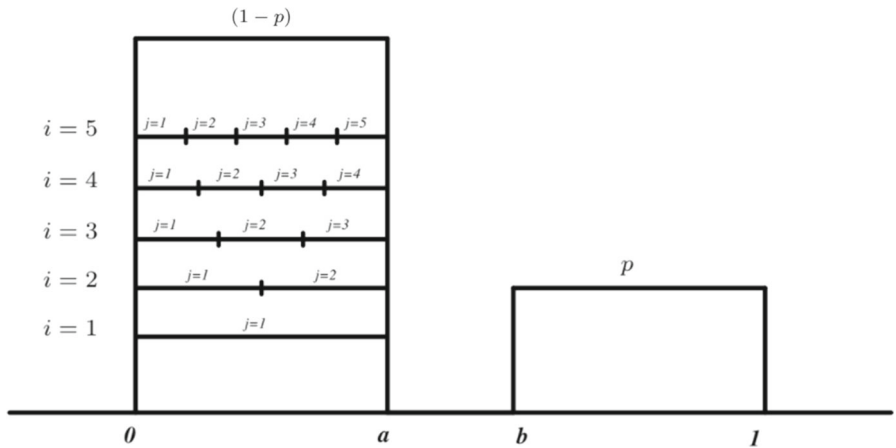

**Fig. 2** Variations in the first mode

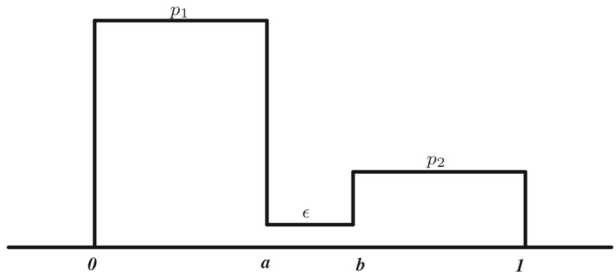

**Fig. 3** Example of a bimodal distribution defined by a convolution of uniforms. Density function

The definition of the 20 scenarios has been carried out symmetrically and with different amplitudes within the original range, just like it is showed in Fig. 3. Specifically, if $a_1$ and $a_2$ are the extremes of the first mode in the definition of the new scenarios, for all different values of $a$, $\forall i = 1,\ldots,5$, $\forall j = 1,\ldots,i$, we define $a_1 = \frac{(j-1)}{i}a$, and $a_2 = \frac{i}{i}a$.

## 5.2 Efficiency of the Draw Auction

In this subsection we test the performance of the Draw auction. So, we compare it with the Myerson auction (Myerson 1981), which is the optimal one. The comparison can be seen in Table 1. Results from Myerson auction and Draw auction have been obtained according to Algorithm 1 *BU Myerson* and Algorithm 2 *Draw Auction*, respectively. Let us remark that, from seller point of view, the Draw auction is nearly as good as Myerson auction. Furthermore, the application of our proposal is immediate, and only two parameters are needed, whereas the definition of Myerson is really complex. In fact, it is not usually applied in real-life sales, except for some particular basic situations. If we focus on the sale of objects as those described in Sect. 2.2, Myerson auction implies a really high complexity level, whereas the application of method proposed in this work, whose results are remarkable, is quite basic.

**Table 1** Efficiency analysis. Draw Auction vs. Myerson Auction

| np | a | b | Draw Auction (%) | Myerson Auction (%) |
|---|---|---|---|---|
| 0.1 | 0.2 | 0.6 | 97.3761 | 100 |
|  |  | 0.8 | 97.1694 | 100 |
|  | 0.4 | 0.6 | 98.8312 | 100 |
|  |  | 0.8 | 97.8979 | 100 |
| 0.2 | 0.2 | 0.6 | 97.6420 | 100 |
|  |  | 0.8 | 97.4181 | 100 |
|  | 0.4 | 0.6 | 98.1642 | 100 |
|  |  | 0.8 | 96.9443 | 100 |
| 0.3 | 0.2 | 0.6 | 97.9499 | 100 |
|  |  | 0.8 | 97.9273 | 100 |
|  | 0.4 | 0.6 | 98.0456 | 100 |
|  |  | 0.8 | 96.8772 | 100 |
| 0.4 | 0.2 | 0.6 | 98.4010 | 100 |
|  |  | 0.8 | 98.3552 | 100 |
|  | 0.4 | 0.6 | 97.9884 | 100 |
|  |  | 0.8 | 97.2911 | 100 |
| 0.5 | 0.2 | 0.6 | 98.5959 | 100 |
|  |  | 0.8 | 98.6227 | 100 |
|  | 0.4 | 0.6 | 98.1828 | 100 |
|  |  | 0.8 | 97.6220 | 100 |

### 5.3 Analysis of Results

In this subsection we compare the Draw auction with three of the more popular auctions, second-price auction without minimum price of sale, second-price auction with minimum price of sale and Myerson auction. We show that the performance of the method proposed in this paper provides better results, from seller viewpoint, than the first two classical auctions, for both original scenario and robustness analysis, and very competitive when comparing it against the optimal model, the Myerson auction (the Draw auction outperforms the Myerson auction in the robustness analysis, and both methods are really similar in the original scenario analysis). Assuming that seller is risk neutral, he chooses the auction with the highest expected profit for himself. In Tables 2 and 3, for two different criteria, we provide the percentage of variation of each auction with regard to the Draw auction, and the parameters of the density function. This percentage of variation about the Draw auction is always 100 in regard to itself.

1. **Original Auction Analysis.** In this case, whose computational results can be observed in Table 2, the Draw and second-price with minimum price of sale auctions are always superior to the second-price auction because they are a generalization of the second-price auction. If we compare the Draw auction with the second-price auction, the first one is, in average, 13.84% better, exceeding it, at most a 34.79% and at least 0.56%. With regard to the second-price auction with minimum price of sale, the Draw auction is 11.81% better in average, a maximum of 27.18% better, and a minimum of 0.56%. Finally, regarding the Myerson auction it is just a 2.18% better in average than the Draw auction, with 3.15% and 1.18% maximum and minimum gain values with respect our proposal.

Table 2 Expected profit for the seller. Original auction analysis

| np | a | b | Second-Price Auction (%) | Second-Price with Minimum Price Auction (%) | Myerson Auction (%) | Draw Auction (%) |
|---|---|---|---|---|---|---|
| 0.1 | 0.2 | 0.6 | 93.7001 | 93.7022 | 102.6947 | 100 |
|  |  | 0.8 | 88.7720 | 88.7740 | 102.9130 | 100 |
|  | 0.4 | 0.6 | 99.4402 | 99.4424 | 101.1827 | 100 |
|  |  | 0.8 | 98.2705 | 98.2728 | 102.1472 | 100 |
| 0.2 | 0.2 | 0.6 | 85.4787 | 85.4807 | 102.4149 | 100 |
|  |  | 0.8 | 77.8016 | 77.8034 | 102.6504 | 100 |
|  | 0.4 | 0.6 | 98.0872 | 98.0895 | 101.8702 | 100 |
|  |  | 0.8 | 95.0018 | 95.0041 | 103.1520 | 100 |
| 0.3 | 0.2 | 0.6 | 79.9258 | 79.9269 | 102.0930 | 100 |
|  |  | 0.8 | 71.0162 | 72.8244 | 102.1166 | 100 |
|  | 0.4 | 0.6 | 96.4602 | 96.4616 | 101.9934 | 100 |
|  |  | 0.8 | 91.6720 | 91.6732 | 103.2235 | 100 |
| 0.4 | 0.2 | 0.6 | 76.2989 | 76.9076 | 101.6250 | 100 |
|  |  | 0.8 | 67.1440 | 79.9276 | 101.6723 | 100 |
|  | 0.4 | 0.6 | 95.1453 | 95.1453 | 102.0529 | 100 |
|  |  | 0.8 | 88.7762 | 88.7762 | 102.7843 | 100 |
| 0.5 | 0.2 | 0.6 | 74.3283 | 79.2970 | 101.4241 | 100 |
|  |  | 0.8 | 65.2090 | 85.6929 | 101.3966 | 100 |
|  | 0.4 | 0.6 | 93.9354 | 93.9361 | 101.8508 | 100 |
|  |  | 0.8 | 86.7159 | 86.7165 | 102.4359 | 100 |

2. **Robustness Analysis.** Results corresponding to this analysis can be seen in Table 3, where, for each value assigned to different parameters of the initial scenario, we show the average of the expected profit for the seller in 20 situations described in previous section.

If we compare the Draw auction with the second-price auction, the first one is, in average, 23.11% better, exceeding it, at most a 46.45%, and, at least, 0.44%. With regard to the second-price auction with minimum price of sale, the Draw auction is 27.62% better in average, a maximum of 46.16%, and a minimum of

Table 3 Expected profit for the seller. Robustness auction analysis

| np | a | b | Second-Price Auction (%) | Second-Price with Minimum Price Auction (%) | Myerson Auction (%) | Draw Auction (%) |
|---|---|---|---|---|---|---|
| 0.1 | 0.2 | 0.6 | 84.6586 | 67.0090 | 70.4813 | 100 |
| | | 0.8 | 78.5162 | 62.2255 | 67.6625 | 100 |
| | 0.4 | 0.6 | 99.5582 | 78.5213 | 79.1336 | 100 |
| | | 0.8 | 98.2965 | 77.5767 | 79.1739 | 100 |
| 0.2 | 0.2 | 0.6 | 72.8402 | 60.6102 | 66.9685 | 100 |
| | | 0.8 | 64.4891 | 53.8419 | 63.4048 | 100 |
| | 0.4 | 0.6 | 93.8982 | 77.3439 | 78.6895 | 100 |
| | | 0.8 | 89.2091 | 73.6203 | 76.8747 | 100 |
| 0.3 | 0.2 | 0.6 | 65.2696 | 56.7093 | 64.9647 | 100 |
| | | 0.8 | 57.1506 | 66.1544 | 62.0763 | 100 |
| | 0.4 | 0.6 | 91.1913 | 78.0094 | 79.9170 | 100 |
| | | 0.8 | 82.2060 | 70.5331 | 74.9767 | 100 |
| 0.4 | 0.2 | 0.6 | 62.5608 | 63.1585 | 65.9477 | 100 |
| | | 0.8 | 54.2953 | 88.1011 | 62.8106 | 100 |
| | 0.4 | 0.6 | 88.0265 | 77.7468 | 79.9875 | 100 |
| | | 0.8 | 78.1570 | 69.2827 | 74.6213 | 100 |
| 0.5 | 0.2 | 0.6 | 61.4374 | 87.3298 | 67.2125 | 100 |
| | | 0.8 | 53.5477 | 92.3400 | 64.9743 | 100 |
| | 0.4 | 0.6 | 86.4071 | 78.3610 | 80.8132 | 100 |
| | | 0.8 | 76.0278 | 69.2169 | 75.2098 | 100 |

7.66%.

Finally, if we pay attention to the Myerson auction, it falls in the robustness analysis with respect to the Draw auction, which is 28.20% better in average, 37.92 at most and 19.19% at least.

In the Tables 2 and 3 it can be seen that, for low values in the first mode (low values of $a$), the Draw auction outperforms other existing methods. Likewise, when it comes to high values in the second mode (high values of $b$), the Draw auction also provides better results than the other auctions. On the other hand, regarding the expected value of the bidders in the second mode (value $np$), the Draw auction reaches the optimal values in the surrounding of 0.4. For values $np \geq 0.5$, the Draw auction loses some effectiveness; however it is still the best auction model for a wide range of combinations of the parameters $a$ and $b$ such that $np \in [0.5, 1]$. When $np > 1$, the improvement of the Draw auction is almost nil.

According to these results and using the criteria which is described above, we can affirm that Draw auction is better than second-price auctions, without or with minimum price of sale, since, in addition to guarantee a higher expected profit for the seller, it is so robust against possible changes in the scenario of application. Moreover, it is really competitive, providing results quite similar to those obtained with the optimal Myerson auction: the Draw auction performs pretty much the same as the Myerson auction in the original scenario, and clearly fits better than the Myerson auction in the robustness analysis.

## 6 Conclusion

In this paper, a new auction, so-called Draw auction, has been introduced. In the Draw auction, a price $c$ is proposed for the $k$ buyers with the highest bids. If more than one buyer is willing to pay the price $c$, then the object is assigned based on the rules of the second-price auction. If only one buyer is willing to pay the price $c$, then the object is assigned to this buyer at a cost of $c$. If none of the buyers is willing to pay the price $c$, then the seller performs a draw among the $k$ buyers, and the winning buyer, who gets the object, pays the amount of the $(k+1)-th$ highest bid.

The Draw auction is applicable in many real-life situations as there are many objects with subjective valuations (as it is shown in Sect. 2.2), which can be modeled as random variables with bimodal density functions. Furthermore, the Draw auction is relatively easy to define, and for each buyer, the corresponding BNE strategy is his personal valuation of the object. We have to take into account that to achieve the BNE is an essential step to analyze an auction.

To test the efficiency of the Draw auction, we compare it with the optimal one, Myerson auction (Myerson 1981). Hence, focusing in the sale of objects whose valuations can be modeled as a bimodal density functions, we show in Table 1 that our proposal is nearly as good as Myerson's one, from seller viewpoint. Moreover, Draw auction's added value is its simplicity. It is a 2-parammetric auction with immediate application, whereas the application of Myerson auction involves a really complex process with.

To carry on with this analysis, it has been necessary to define explicitly Myerson auction for the case of a convolution of uniform distributions in which the first mode is much greater than the second one. Thus, we have defined the Algorithm *BU Myerson* to calculate, with complexity $O(n)$, the value of Myerson auction for the case of bimodal distributions. It is important to take into account that this explicit definition is something new in the literature.

Another objective of this paper has been to compare the expected profit for the seller in the Draw auction, for the case of bimodal distributions, with regard to two of the most popular auctions: second-price, without and with minimum price of sale. For doing it, two possible criteria have been considered, which are detailed in Sect. 5.3, namely, Original Auction Analysis, and Robustness Analysis.

The analysis of original auction allows us to affirm that Draw auction provides an optimal expected profit for the seller, when comparing it with the more used auctions, second-price auctions, with and without minimum price of sale. Then, in robustness analysis, Draw auction is the procedure which guarantees an optimal expected profit.

In conclusion, we define an auction, known as the draw auction, which is suitable for objects with subjective valuation that can be modeled as random variables with bimodal distribution, in which the first mode is much greater than the second one (under the assumptions of the benchmark model and considering model's robustness). The draw auction exhibits superior behavior compared to other existing auctions, as second-price auction, or second-price with minimum price of sale auction. Furthermore, the Draw auction performs nearly as good as Myerson auction. In addition, the Draw auction employs no a minimum price, so the object is always sold. Therefore, for the application in real-life in the auction of objects whose valuation can be modeled as a bimodal distribution in which the first mode is greater than the second one, Draw auction should be used, as, with a really simple definition and application, it provides a high expected profit for the seller (near to the optimal one), in addition to ensuring the necessary robustness against possible changes in the scenario when it is executed.

## Appendix A: Notation

In this section we provide a list with the notation used in order of appearance.

- $N = \{1, \ldots, n\}$: set of players.
- $S_i$: player $i$'s action or strategy space.
- $(s_1, \ldots, s_n)$: set of actions chosen by a player.
- $w_i$: player $i$'s type.
- $W_i$: set of possible types related to the player $i$.
- $w_i \in W_i$: player $i$'s type, privately known by himself.
- $u_i(s_1, \ldots, s_n; w_i)$: player $i$'s payoff when the $i$ choose the actions $(s_1, \ldots, s_n)$, and $w_i \in W_i$ is player $i$'s type.
- $\mathcal{G} = \{S_1, \ldots, S_n; W_1, \ldots, W_n; p_1, \ldots, p_n; u_1, \ldots, u_n\}$: Bayesian Game.
- BNE: Bayesian Nash Equilibrium.
- $p_i(w_{-i} \mid w_i)$: uncertainty about the $n-1$ other players' possible types, $w_{-i}$, given his own type, $w_i$.
- $(p_1, \ldots, p_n)$: players' beliefs, when the belief of player $i$ is $p_i(w_{-i} \mid w_i)$.
- $s_i(w_i)$: strategy for player $i$, the action from the feasible set $S_i$ that type $w_i$ would choose if drawn by nature
- $E_i[s_1, \ldots, s_n] = \sum_{w_{-i} \in W_{-i}} u_i(s_1, \ldots, s_n; v) p_i(w_{-i} \mid w_i)$: expected value of player $i$.
- $F$: distribution function related to a given random variable.
- $F^{-1}$: inverse function of $F$.
- $f$: density function related to a given random variable.
- $H$: auxiliary function used by Myerson.
- $G$: convex hull of the function $H$.
- $g(q) = G'(q)$.
- $M(x)$: auxiliary set in the original Myerson auction definition.
- $\bar{c}(x_i)$: auxiliary function in the original Myerson auction definition.
- $\bar{p}_i(x)$: probability that the object is assigned to player $i$.

- $\bar{p}(x) = (\bar{p}_1(x), \ldots, \bar{p}_n(x))$: vector of probabilities that the object is assigned to a bidder.
- $\bar{x}_i(x)$: expected payment of the object done by player $i$.
- $\bar{x}(x) = (\bar{x}_1(x), \ldots, \bar{x}_n(x))$: vector of expected payments of the object for each player.
- $p$: probability of having a 'high' valuation.
- $k$: amount of 'high' valuations.
- $(1 - p)$: probability of having a 'low' valuation.
- $n - k$: amount of 'low' valuations.
- $p_1$: cumulative probability of the first mode.
- $p_2$: cumulative probability of the second mode.
- $a$: top limit of the first mode.
- $b$: lower limit of the second mode.
- $\epsilon$: cumulative probability between both modes.
- $x_i$: bid offered by the player $i$ for buying the object.
- $x = (x_1, \ldots, x_n)$: vector of buyers bids.
- $(x_{(1)}, \ldots, x_{(n)})$: ordered list of bids, where $x_{(1)}$ is the highest bid, $x_{(2)}$ is the second highest bid, and so on.
- $x_{min} = \frac{a}{2p_1}$: first minimum price of sale.
- $x_{\ell\ell} = \frac{az_0}{p_1}$: lower limit of the Myerson draw.
- $x_{cut}$: upper limit of the Myerson draw or second minimum price of sale.
- $\beta_0$: bound needed to fix the minimum sale value, . If $\beta_0 < 0$, the minimum selling price is the upper limit of Myerson draw. Otherwise, this minimum price is the first minimum price of sale, $x_{min}$.
- $\mathcal{D}$: set of buyers whose valuation belongs to the draw interval in the *BU Myerson* Algorithm.
- $A$: auxiliary point used in *BU Myerson* Algorithm.
- $B$: auxiliary point used in *BU Myerson* Algorithm.
- $C$: auxiliary point used in *BU Myerson* Algorithm.
- $D$: auxiliary point used in *BU Myerson* Algorithm.
- $E$: auxiliary point used in *BU Myerson* Algorithm.
- $y_1$: positive root of the equation $Cx^2 + Dx + E = 0$.
- $y_2$: negative root of the equation $Cx^2 + Dx + E = 0$.
- $z_1$: auxiliary point used in *BU Myerson* Algorithm.
- $z_2$: auxiliary point used in *BU Myerson* Algorithm.
- $z_3$: auxiliary point used in *BU Myerson* Algorithm.
- $H_1$: first part of the function $H$, defined in $[0, p_1]$
- $H_2$: second part of the function $H$, defined in $(p_1, p_1 + \epsilon]$
- $H_3$: third part of the function $H$, defined in $(p_1 + \epsilon, 1]$
- $z_0$: auxiliary point used in *BU Myerson* Algorithm.
- $y_0$: auxiliary point used in *BU Myerson* Algorithm.
- $t_1$: tangent line to $H_1$ at the point $s_1$.
- $t_3$: tangent line to $H_3$ at the point $s_3$.
- $r_1$: line through the points $(z_1, H_1(z_1))$ and $(y_1, H_3(y_1))$.
- $r_2$: line through the points $(z_2, H_1(z_1))$ and $(y_2, H_3(y_2))$.

- $r_3$: tangent line to $H_1$ which goes through the point $((p_1 + \epsilon), H_3(p_1 + \epsilon))$.
- $r_4$: line through the points $(0, 0)$ and $((p_1 + \epsilon), H_3(p_1 + \epsilon))$
- $\alpha_i$: independent term of the line $r_i$.
- $\beta_i$: slope of the line $r_i$.
- $\alpha_0$: auxiliary point used in the proof of *BU Myerson* Algorithm.
- $v_0$: personal valuation of the object given by the seller.
- $v_i$: valuation of the object given by player $i \in N$.
- $v = (v_1, \ldots, v_n)$: vector of buyers valuations.
- $\left(v_{(1)}, \ldots, v_{(n)}\right)$ denotes the ordered vector of valuations, where $v_{(1)}$ is the highest valuation, $v_{(2)}$ is the second highest valuation, and so on.
- $c > v_0$: constant value which represents a fixed price.
- $x^* = (x_1, \ldots, x_{i-1}, v_i, x_{i+1}, \ldots, x_n)$: modified bid vector.
- $\Re_+^n$: set of $n$-vectors whose elements are real not negative numbers.
- $f_1$: density function related to having a high probability of assigning a low valuation.
- $f_2$: density function related to having a low probability of assigning a high valuation.
- $v^*$: minimum price fixed in the second-price auction with minimum price.
- $Z$: random variable which represents the number of valuations that satisfy a particular condition.
- $W$: random variable which represents the number of valuations that satisfy a particular condition.
- $D_k$: set of buyers with the $k$ highest bids in the Draw auction.

## Appendix B: Proofs

In this section we provide the demonstration of the Proposition 1 and Proposition 2, in order of appearance.

**Proof** Here we provide the proof of the Proposition 1. We give a detailed step-by-step explanation. We consider the density function $f$ (see Fig. 3), where

$$f(x) = \begin{cases} \dfrac{p_1}{a}, & \text{if } x \in (0, a] \\ \dfrac{\epsilon}{b - a}, & \text{if } x \in (a, b] \\ \dfrac{p_2}{1 - b}, & \text{if } x \in (b, 1] \end{cases}$$

The related distribution function of $f$ is $F$, where

$$F(x) = \begin{cases} 0, & \text{if } x \leq 0 \\ \dfrac{p_1}{a}x, & \text{if } x \in (0, a] \\ p_1 + \dfrac{\epsilon(x-a)}{b-a}, & \text{if } x \in (a, b] \\ p_1 + \epsilon + \dfrac{p_2(x-b)}{1-b}, & \text{if } x \in (b, 1] \end{cases}$$

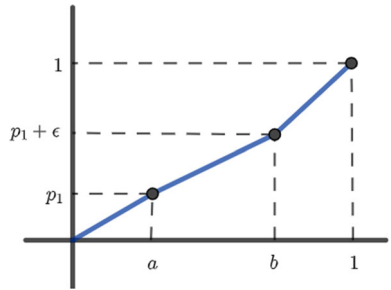

**Fig. 4** Function $F^{-1}$, inverse function of the distribution $F$

According to Myerson's notation, let $H(q) = \int_0^q h(r)dr$, where the function $h$ is defined as $h(q) = F^{-1}(q) - \dfrac{1-q}{f(F^{-1}(q))}$.

Then, to define the function $H$, we have to calculate the inverse function of $F$, $F^{-1}$, shown in the Fig. 4, where

$$F^{-1}(x) = \begin{cases} \dfrac{ax}{p_1}, & \text{if } x \in (0, p_1] \\ a + \dfrac{(b-a)(x-p_1)}{\epsilon}, & \text{if } x \in (p_1, p_1+\epsilon] \\ b + \dfrac{(1-b)(x-p_1-\epsilon)}{p_2}, & \text{if } x \in (p_1+\epsilon, 1) \end{cases}$$

Then, we have to calculate the value of $f$ applied in $F^{-1}$, so that

$$f(F^{-1}(x)) = \begin{cases} \dfrac{p_1}{a}, & \text{if } x \in (0, p_1] \\ \dfrac{\epsilon}{b-a}, & \text{if } x \in (p_1, p_1+\epsilon] \\ \dfrac{p_2}{1-b}, & \text{if } x \in (p_1+\epsilon, 1) \end{cases}$$

At this point, it is quite simple to obtain the function $h(q) = F^{-1}(q) - \dfrac{1-q}{f(F^{-1}(q))}$, where

$$f(x) = \begin{cases} \dfrac{ax}{p_1} - \dfrac{1-x}{\frac{p_1}{a}} = \dfrac{a}{p_1}(2x-1), & \text{if } x \in (0, p_1] \\ a + \dfrac{(b-a)(x-p_1)}{\epsilon} - \dfrac{1-x}{\frac{\epsilon}{b-a}} = a + \dfrac{b-a}{\epsilon}(2x - (1-p_1)), & \text{if } x \in (p_1, p_1 + \epsilon] \\ b + \dfrac{(1-b)(x-p_1-\epsilon)}{p_2} - \dfrac{1-x}{\frac{p_2}{1-b}} = b + \dfrac{1-b}{p_2}(2x - (1+p_1+\epsilon)), & \text{if } x \in (p_1+\epsilon, 1) \end{cases}$$

Then, $H(q) = \int_0^q h(r)dr$ is a piecewise function characterized as follows:

$$H(q) = \begin{cases} H_1(q) = \dfrac{a}{p_1}(q^2 - q) & \text{if } q \in [0, p_1] \\ H_2(q) = a(q-1) + \\ \dfrac{(b-a)}{\epsilon}(q^2 - (1+p_1)q + p_1) & \text{if } q \in (p_1, p_1+\epsilon] \\ H_3(q) = b(q-1) + \dfrac{(1-b)}{p_2}(q^2 - (1+p_1+\epsilon)q \\ + (p_1+\epsilon)) & \text{if } q \in (p_1+\epsilon, 1] \end{cases} \tag{B1}$$

The demonstration we are facing is based on the calculation of the convex hull of $H$, called $G$.

$H$ is a continuous function, as $H_1(p_1) = \lim_{q \to p_1} H_2(q)$ and $H_2(p_1 + \epsilon) = \lim_{q \to p_1 + \epsilon} H_3(q)$. Moreover, there is a lack of convexity between $H_2$ and $H_3$, as $H_1$ and $H_2$ are decreasing functions and $H_3$ is increasing in the specific bimodal distributions we have considered. Due to this lack of convexity, we have to connect some points of $H_1$ with any other of $H_3$ abide the convexity, in order to calculate the convex hull. This connection does not arise regarding $H_2$, as we need to calculate the limit when $\epsilon \to 0$, scenario in which $H_2$ does not exist.

To obtain the relative connections of $H_1$ and $H_3$, we work with the definition of a tangent line between those functions, under the assumption that the points of tangency have to be in the definition interval of $H_1$ and $H_3$, which are $[0, p_1]$ and $(p_1 + \epsilon, 1]$ respectively. The calculation of this tangent line are showed below, in the point 1.

If we can not obtain that tangent line to $H_1$ and $H_3$, the convex hull is calculated considering a point of tangency to one of them, and the extreme of the other. In other words, we calculate a tangent line to $H_1$ which passes through the point $((p_1 + \epsilon), H_3(p_1 + \epsilon))$ (the extreme point of $H_3$), or we calculate the tangent line to $H_3$ which passes through the point $(0, 0)$ (the extreme point of $H_1$). However, as $H_1$ and $H_2$ are decreasing, and $H_3$ is increasing, this scenario is not feasible, so we only have to consider the other option. We show the calculation of that line below in the point 2.

Finally, if the tangent line to $H_1$ can not be calculated with point of tangency in its definition interval $[0, p_1]$, the convex hull will be calculated by considering a line which connects the points $(0, 0)$ and $((p_1 + \epsilon), H_3(p_1 + \epsilon))$. We show the calculation of that line below in the point 3.

1. The tangent lines among $H_1$ and $H_3$.
   Let $t_1(s)$ be the tangent line to $H_1$ at the point $s_1$, $t_1(s) = \frac{a}{p_1}(2s_1 - 1)(s - s_1) + \frac{a}{p_1}(s_1^2 - s_1)$; and $t_3(s)$ the tangent line to $H_3$ at the point $s_3$, $t_3(s) = [b + \frac{1-b}{p_2}(2s_3 - (1+p_1+\epsilon))](s - s_3) + b(s_3 - 1) + \frac{1-b}{p_2}(s_3^2 - (1+p_1+\epsilon)s_3 + (p_1+\epsilon))$.
   To calculate the tangent line to curves $H_1$ and $H_3$, $t_1$ and $t_3$ must match both on the slope and in the independent terms. If we equalize these terms, we get the values mentioned in the Algorithm *BU Myerson*, $z_1, y_1$, which will be used to obtain the first tangent line, $r_1$; and $z_2, y_2$, which are needed to define the second tangent line, $r_2$.

- The first line, $r_1$, will pass through the points $(z_1, H_1(z_1))$ and $(y_1, H_3(y_1))$.

$$r_1(q) := H_1(z_1)\left(1 - \frac{H_3(y_1) - H_1(z_1)}{y_1 - z_1}\right) + \left(\frac{H_3(y_1) - H_1(z_1)}{y_1 - z_1}\right)q \quad \text{(B2)}$$

- The second one, $r_2$, will pass through the points $(z_2, H_1(z_2))$ y $(y_2, H_3(y_2))$.

$$r_2(q) := H_1(z_2)\left(1 - \frac{H_3(y_2) - H_1(z_2)}{y_2 - z_2}\right) + \left(\frac{H_3(y_2) - H_1(z_2)}{y_2 - z_2}\right)q \quad \text{(B3)}$$

2. The tangent line to $H_1$ which passes through the point $((p_1 + \epsilon), H_3(p_1 + \epsilon))$. Substituting this point, $((p_1 + \epsilon), H_3(p_1 + \epsilon))$, in the equation of $t_1(s)$, we get its tangency point in $H_1$,

$$z_3 = (p_1 + \epsilon) - \sqrt{(p_1 + \epsilon)^2 - (p_1 + \epsilon) - \frac{p_1 b}{a}(p_1 + \epsilon - 1)}. \quad \text{(B4)}$$

Then, the expected line will be $r_3$:

$$r_3(q) := H_3(z_3)\left(1 - \frac{H_3(p_1 + \epsilon) - H_3(z_3)}{(p_1 + \epsilon) - z_3}\right) + \left(\frac{H_3(p_1 + \epsilon) - H_3(z_3)}{(p_1 + \epsilon) - z_3}\right)q \quad \text{(B5)}$$

3. The line that passes through the points $(0, 0)$ and $((p_1 + \epsilon), H_3(p_1 + \epsilon))$.

$$r_4(q) := \left(\frac{H_3(p_1 + \epsilon)}{p_1 + \epsilon}\right)q \quad \text{(B6)}$$

Once all the lines have been defined, we have to work with their components. Then, for each line $r_i$, let $\alpha_i$ be its independent term, and $\beta_i$ its slope, so that:

- $\alpha_1 = H_1(z_1)\left(1 - \frac{H_3(y_1) - H_1(z_1)}{y_1 - z_1}\right)$ and $\beta_1 = \frac{H_3(y_1) - H_1(z_1)}{y_1 - z_1}$
- $\alpha_2 = H_1(z_2)\left(1 - \frac{H_3(y_2) - H_1(z_2)}{y_2 - z_2}\right)$ and $\beta_2 = \frac{H_3(y_2) - H_1(z_2)}{y_2 - z_2}$
- $\alpha_3 = H_3(z_3)\left(1 - \frac{H_3(p_1+\epsilon) - H_3(z_3)}{(p_1+\epsilon) - z_3}\right)$ and $\beta_3 = \frac{H_3(p_1+\epsilon) - H_3(z_3)}{(p_1+\epsilon) - z_3}$
- $\alpha_4 = 0$ and $\beta_4 = \frac{H_3(p_1+\epsilon)}{p_1+\epsilon}$

To obtain the convex hull, we also need to define points $z_0, y_0, \alpha_0, \beta_0$:

- If $z_1 \in [0, p_1] \wedge y_1 \in [p_1 + \epsilon, 1] \implies z_0 := z_1, \ y_0 := y_1, \ \alpha_0 := \alpha_1, \ \beta_0 := \beta_1$

- If $z_2 \in [0, p_1] \wedge y_2 \in [p_1 + \epsilon, 1] \implies z_0 := z_2, y_0 := y_2, \alpha_0 := \alpha_2, \beta_0 := \beta_2$
- If $z_3 \in [0, p_1] \implies z_0 := z_3, y_0 := (p_1 + \epsilon), \alpha_0 := \alpha_3, \beta_0 := \beta_3$
- Otherwise $z_0 := 0, y_0 := (p_1 + \epsilon), \alpha_0 := \alpha_4, \beta_0 := \beta_4$

Therefore, convex hull will be formalized as:

$$G(q) = \begin{cases} H_1(q) & \text{if } q \in [0, z_0] \\ \alpha_0 + \beta_0 q & \text{if } q \in (z_0, y_0] \\ H_3(q) & \text{if } q \in (y_0, 1] \end{cases} \quad (B7)$$

According to Myerson's notation, the following holds:

- $g(q) := G'(q)$
- $\bar{c}(x_i) := g(F(x_i))$
- $M(x) := \{i \mid 0 \leq \bar{c}(x_i) = max_{j \in N}(\bar{c}(x_j))\}$
- The probability that the object is assigned to the bidder $i$ is:
  - $\bar{p}_i(x) = \frac{1}{|M(x)|}$, if $i \in M(x)$
  - 0, otherwise

- The function that defines the expected payment of the object is $\bar{x}_i(x) = \bar{p}_i(x)x_i - \int_{a_i}^{x_i} \bar{p}_i(x_{-i}, s_i) ds_i$.

Then, we have

$$g(q) = G'(q) = \begin{cases} \dfrac{a}{p_1}(2q - 1) & \text{if } q \in [0, z_0] \\ \beta_0 & \text{if } q \in (z_0, y_0] \\ b + \dfrac{(1-b)}{p_2}(2q - (1 + p_1 + \epsilon)) & \text{if } q \in (y_0, 1] \end{cases} \quad (B8)$$

$$\bar{c}(x_i) = g(F(x_i)) = \begin{cases} \dfrac{a}{p_1}(2F(x_i) - 1) = \dfrac{a}{p_1}\left(2\dfrac{p_1}{a}x_i - 1\right) & \text{if } F(x_i) \in [0, z_0] \\ \beta_0 & \text{if } F(x_i) \in (z_0, y_0] \\ b + \dfrac{(1-b)}{p_2}(2F(x_i) - (1 + p_1 + \epsilon)) = \\ b + \dfrac{(1-b)}{p_2}\left(2\left(p_1 + \epsilon + \dfrac{p_2(x_i - b)}{1 - b}\right) - (1 + p_1 + \epsilon)\right) & \text{if } F(x_i) \in (y_0, 1] \end{cases}$$
$$(B9)$$

To continue with the calculation of $M(x)$, we have to distinguish different cases depending on whether $\beta_0$ is positive or negative and on the bids of the different players. In those scenario we consider the notation $x_{cut} = F^{-1}(y_0)$, $x_{\ell\ell} = F^{-1}(z_0) = \frac{az_0}{p_1}$ if $z_0 \in [0, p_1]$, and $x_{min} = \frac{a}{2p_1}$.

1. If $\beta_0 < 0$, as $F(x_i)$ is an increasing function, $\bar{c}(x_i) \geq 0$ only for values $x_i \in [F^{-1}(y_0), 1] = [x_{cut}, 1]$.

   (1a) If $x_{(1)} \in [x_{cut}, 1]$ and $x_{(2)} \notin [x_{cut}, 1]$, then $M(x) = \{(1)\}$, $\bar{p}_{(1)}(x) = 1$ and $\bar{x}_{(1)}(x) = 1x_{(1)} - (x_{(1)} - x_{cut}) = x_{cut}$. It is, the player who offers the higher bid gets the object, paying $x_{cut}$ for it.

   (1b) If $x_{(1)}, x_{(2)} \in [x_{cut}, 1]$, then $M(x) = \{(1)\}$, $\bar{p}_{(1)}(x) = 1$ and $\bar{x}_{(1)}(x) = 1x_{(1)} - (x_{(1)} - x_{(2)}) = x_{(2)}$. It is, the player who offers the higher bid gets the object, paying $x_{(2)}$ for it.

   (1c) If $x_{(1)} \notin [x_{cut}, 1]$, then $M(x) = \emptyset$, $\bar{p}_i(x) = 0 \; \forall i \in \{1, \ldots, n\}$ and $\bar{x}_i(x) = 0 \; \forall i \in \{1, \ldots, n\}$. In this case the object is not sold.

2. If $\beta_0 \geq 0$, as $F(x_i)$ is an increasing function it holds that $\bar{c}(x_i) \geq 0$ if $\frac{a}{p_1}\left(2\frac{p_1}{a}x_i - 1\right) \geq 0 \implies x_i \geq \frac{a}{2p_1} = x_{min}$.

   (2a) If $x_{(1)} \geq x_{min}$ and $x_{(2)} < x_{min}$, then $M(x) = \{(1)\}$, $\bar{p}_{(1)}(x) = 1$ and $\bar{x}_{(1)}(x) = 1x_{(1)} - (x_{(1)} - x_{min}) = x_{min}$. It is, the player who offers the higher bid gets the object, paying $x_{min}$ for it.

   (2b) If $x_{(1)}, x_{(2)} \in [x_{min}, x_{\ell\ell}]$, then $M(x) = \{(1)\}$, $\bar{p}_{(1)}(x) = 1$ and $\bar{x}_{(1)}(x) = 1x_{(1)} - (x_{(1)} - x_{(2)}) = x_{(2)}$. It is, the player who offers the higher bid gets the object, paying $x_{(2)}$ for it.

   (2c) If $x_{(1)} \in (x_{\ell\ell}, x_{cut}]$ and $x_{(2)} \in [x_{min}, x_{\ell\ell})$, then $M(x) = \{(1)\}$, $\bar{p}_{(1)} = 1$, and $\bar{x}_{(1)}(x) = 1x_{(1)} - (x_{(1)} - x_{(2)}) = x_{(2)}$. It is, the player who offers the higher bid gets the object, paying $x_{(2)}$ for it.

   (2d) If $x_{(1)}, x_{(2)} \in (x_{\ell\ell}, x_{cut}]$, then $M(x) = \{(i) \mid x_{(i)} \in (x_{\ell\ell}, x_{cut}]\}$, $\bar{p}_{(i)}(x) = \frac{1}{|M(x)|}$, and $\bar{x}_{(i)}(x) = \frac{1}{|M(x)|}(x_{(i)} - (x_{(i)} - x_{\ell\ell})) = \frac{x_{\ell\ell}}{|M(x)|}$, $\forall i \in M(x)$. In this case the object is raffled among all the bidders in $M(x)$; the winner of the draw obtains the object, paying $x_{\ell\ell}$ for it (before the draw, the expected value of payment of each player is $\frac{x_{\ell\ell}}{|M(x)|}$).

   (2e) If $x_{(1)} > x_{cut}$ and $x_{(2)} \in (x_{\ell\ell}, x_{cut}]$, then $M(x) = \{(1)\}$, $\bar{p}_{(1)}(x) = 1$, and $\bar{x}_{(1)}(x) = 1x_{(1)} - \left(x_{(1)} - \left(x_{cut} - \frac{x_{cut} - x_{\ell\ell}}{|D|+1}\right)\right) = x_{cut} - \frac{x_{cut} - x_{\ell\ell}}{|D|+1}$, being $D = \{(i) \mid x_{(i)} \in (x_{\ell\ell}, x_{cut}]\}$. In this case, the player who offers the higher bid gets the object, paying for it the minimum value he can pay without compensation to enter the draw.

   (2f) If $x_{(1)} > x_{cut}$, $x_{(2)} \geq x_{min}$ and $x_{(2)} \in (x_{\ell\ell, x_{cut}}]$, then $M(x) = \{(1)\}$, $\bar{p}_{(1)}(x) = 1$ and $\bar{x}_{(1)}(x) = 1x_{(1)} - (x_{(1)} - x_{(2)}) = x_{(2)}$. It is, the player who offers the higher bid gets the object, paying $x_{(2)}$ for it.

   (2g) If $x_{(1)} < x_{min}$, then $M(x) = \emptyset$, $\bar{p}_i(x) = 0 \; \forall i \in \{1, \ldots, n\}$, and $\bar{x}_i(x) = 0 \; \forall i \in \{1, \ldots, n\}$. In this case the object is not sold.

This casuistry is performed by the algorithm BU Myerson, with the only difference that some conditions are omitted in the algorithm, as for example (2f). The algorithm skips the conditions $x_{(2)} \geq x_{min}$ and $x_{(2)} \in (x_{\ell\ell}, x_{cut}]$ as, at this point of the *if/else* sequence, the rest of feasible values for $x_{(2)}$ have already been ruled out. □

**Proof** Here we provide the proof of the Proposition 2.

Generally speaking, it is complicated to prove that demonstration for any bids vector $x$, as there are very diverse possibilities. To develop this demonstration, we do a partition of the family of the different types of bid vectors, so, in each group, the vectors considered have several characteristics in common which allow us to easily demonstrate that Proposition. Then, we will carry out a process of demonstration by exhaustion. It will be based on the answer of several dichotomous questions. The demonstration is long, because of the wide variety of different cases, but it is really easy in all of them.

We start from the affirmation which asserts that every dominant strategy is also a BNE strategy. Then, we only have to demonstrate that the bid vector composed by buyers' valuations is a dominant strategy. Formally, for all bid vectors, $x = (x_1, \ldots, x_n)$, where $x_i \neq v_i$, the vector $x^* = (x_1, \ldots, x_{i-1}, v_i, x_{i+1}, \ldots, x_n)$ has an expected profit for the $i$-th buyer that is greater than or equal to the expected profit for $x$, i.e. $E_i[x^*] \geq E_i[x] \ \forall x \in \mathfrak{R}_+^n$. Let $x_{(j)}$ be the $j$-th highest bid in $x$ and let $x_{(j)}^*$ denote the $j$-th highest bid in $x^*$. The comparison between $E_i[x^*]$ and $E_i[x]$ will be performed by considering several feasible cases. To define these cases, we consider the following eight Yes/Not questions:

**$Q_1$**: $x_i > x_{(k+1)}$? (Is the $i$-th buyer among the $k$ buyers with the highest bids in $x$?).

**$Q_2$**: $v_i > x_{(k+1)}^*$? (Is the $i$-th buyer among the $k$ buyers with the highest bids in $x^*$?).

**$Q_3$**: $x_i > x_{(k+1)}$ and $x_i - c > \frac{1}{k}(x_i - x_{(k+1)})$? (According to $x$, is the $i$-th buyer interested in paying the price $c$ for the object, and can he afford it?).

**$Q_4$**: $v_i > x_{(k+1)}^*$ and $v_i - c > \frac{1}{k}(v_i - x_{(k+1)}^*)$? (According to $x^*$, is the $i$-th buyer interested in paying the price $c$ for the object, and can he afford it?).

**$Q_5$**:
$\exists j \in \{1, \ldots, i-1, i+1, \ldots, n\}$ such that $x_j > x_{(k+1)}$ and $x_j - c > \frac{1}{k}(x_j - x_{(k+1)})$? (According to $x$, are there any other buyers, different from the $i$-th buyer, who can afford the price $c$ and who interested in paying this price for the object?).

**$Q_6$**:
$\exists j \in \{1, \ldots, i-1, i+1, \ldots, n\}$ such that $x_j > x_{(k+1)}^*$ and $x_j - c > \frac{1}{k}(x_j - x_{(k+1)}^*)$? (According to $x^*$, are there any other buyers, different from the $i$-th buyer, who can afford the price $c$ and who interested in paying this price for the object?)

**$Q_7$**: $x_i > x_{(2)}$? (Does the $i$-th buyer have the highest bid in $x$?)

**$Q_8$**: $v_i > x_{(2)}^*$? (Does the $i$-th buyer have the highest bid in $x^*$?)

In addition to the $2^8 = 256$ possible cases determined by the previous questions, we make an initial distinction depending on whether $x_i < v_i$ or $x_i > v_i$. For $x_i < v_i$, we will take into account the following relations between the questions to reduce the number of cases, assuming that $Q_i$ denotes an affirmative answer to the $i$-th question and $\overline{Q}_i$ denotes a negative answer to the $i$-th question.

($I_1$) $\overline{Q}_1 \to \overline{Q}_3$ and $\overline{Q}_7$: if the $i$-th buyer is not one of the $k$ highest buyers, then he can not afford the price $c$ and is not the highest buyer.

($I_2$) $\overline{Q}_2 \to \overline{Q}_4$ and $\overline{Q}_8$: if the $i$-th buyer is not one of the $k$ highest buyers, then he can not afford the price $c$ and is not the highest buyer.

($I_3$) $\overline{Q}_3$ and $Q_5 \to \overline{Q}_7$: if the $i$-th buyer is not interested in paying $c$ and there is another buyer who is interested in paying $c$, then the $i$-th buyer is not the highest buyer.

($I_4$) $\overline{Q}_4$ and $Q_6 \to \overline{Q}_8$: if the $i$-th buyer is not interested in paying $c$ and there is another buyer who is interested in paying $c$, then the $i$-th buyer is not the highest buyer.

($I_5$) $Q_3$ and $\overline{Q}_5 \to Q_7$: if the $i$-th buyer is interested in paying $c$, and no other buyer is interested in paying $c$, then the $i$-th buyer places the maximum bid.

($I_6$) $Q_4$ and $\overline{Q}_6 \to Q_8$: if the $i$-th buyer is interested in paying $c$, and no other buyer is interested in paying $c$, then the $i$-th buyer places the maximum bid.

($I_7$) $Q_1$ and $Q_2 \to (Q_5$ and $Q_6)$ or $(\overline{Q}_5$ and $\overline{Q}_6)$: if in both bid vectors, $x$ and $x^*$, the $i$-th buyer places the $k$-th highest bid, then $x_{(k+1)} = x^*_{(k+1)}$. Therefore, the questions $Q_5$ and $Q_6$ will have the same answer.

($I_8$) $Q_1 \to Q_2$: given that $x_i < v_i$: if the $i$-th bid in $x$ is among the $k$ highest bids, then the $i$-th bid in $x^*$ is also one of the $k$ highest bids.

($I_9$) $Q_3 \to Q_4$: given that $x_i < v_i$, if the $i$-th bid in $x$ indicates that the $i$-th buyer is interested in paying $c$, then the $i$-th bid in $x^*$ also indicates that he is interested in paying $c$.

($I_{10}$) $Q_7 \to Q_8$: given that $x_i < v_i$, if the $i$-th bid in $x$ is the highest, then the $i$-th bid in $x^*$ will also be the highest one.

Considering these 10 relations, and grouping similar cases, we reduce the 256 original cases to 15 final cases. Figure 5 illustrates the process. When a node is not divided into two, the implication ($I_i$) or a question mark (?) appears. When the implication $I_i$ is indicated, there is only one possibility. The question mark indicates that both possibilities are included in the same case.

For these 15 final cases, we calculate the expected profit for the $i$-th buyer considering both bid vectors, $x$ and $x^*$, and we compare the respective values:

- **Case 1:** $\overline{Q}_1$ for $x$ and $\overline{Q}_2$ for $x^*$, then, $E_i[x] = E_i[x^*] = 0$ (the $i$-th buyer is not among the $k$ highest buyers in either $x$ or $x^*$).
- **Case 2:** $\overline{Q}_1$, so that $E_i[x] = 0$ (the $i$-th buyer is not among the $k$ highest buyers in $x$). Moreover, $Q_2$, $\overline{Q}_4$ and $\overline{Q}_6$ (the $i$-th buyer is among the $k$ highest buyers in $x^*$, however he is not interested in paying $c$ and no other buyer is interested in paying $c$). Then, as $v_i > x^*_{(k+1)}$, $E_i[x^*] = \frac{1}{k}\left(v_i - x^*_{(k+1)}\right)$. Due to $Q_2$, $E_i[x^*] = \frac{1}{k}\left(v_i - x^*_{(k+1)}\right) > 0 = E_i[x]$.

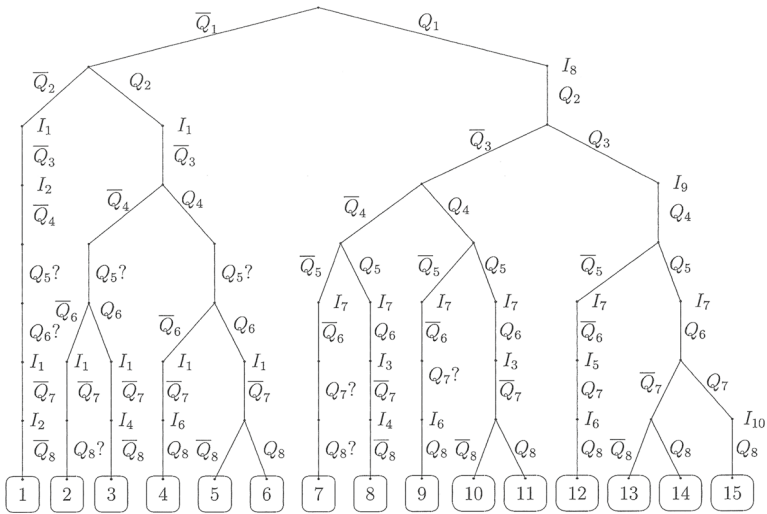

**Fig. 5** Process of reducing the 256 original cases to the 15 final cases when $x_i < v_i$

- **Case 3:** $\overline{Q}_1$, so $E_i[x] = 0$ (the $i$-th buyer is not among the $k$ highest buyers in $x$). Also, $Q_2$, $\overline{Q}_4$ and $Q_6$, so $E_i[x^*] = 0$ (the $i$-th buyer is among the $k$ highest buyers in $x^*$, however he is not interested in paying $c$ and someone else is interested in paying $c$).

- **Case 4:** $\overline{Q}_1$, so $E_i[x] = 0$ (the $i$-th buyer is not among the $k$ highest buyers in $x$). Also, $Q_2$, $Q_4$ and $\overline{Q}_6$, then, $E_i[x^*] = v_i - c$ (the $i$-th buyer is among the $k$ highest buyers in $x^*$, and he is interested in paying $c$, while no other buyer is interested in paying $c$). Because $v_i > x^*_{(k+1)}$ and $v_i - c > \frac{1}{k}\left(v_i - x^*_{(k+1)}\right)$, due to $Q_4$, we have $E_i[x^*] = v_i - c_i > \frac{1}{k}\left(v_i - x^*_{(k+1)}\right) > 0 = E_i[x]$.

- **Case 5:** $\overline{Q}_1$, so $E_i[x] = 0$ (the $i$-th buyer is not among the $k$ highest buyers in $x$). Also, $Q_2$, $Q_4$, $Q_6$ and $\overline{Q}_8$, so $E_i[x^*] = 0$ (the $i$-th buyer is among the $k$ highest buyers in $x^*$, and both, he and others, are interested in paying $c$; however he is not the highest buyer).

- **Case 6:** $\overline{Q}_1$, so $E_i[x] = 0$ (the $i$-th buyer is not among the $k$ highest buyers in $x$). Also $Q_2$, $Q_4$, $Q_6$ and $Q_8$, then, $E_i[x^*] = v_i - x^*_{(2)}$ (the $i$-th buyer is among the $k$ highest buyers in $x^*$, and both, he and others are interested in paying $c$; however he is the highest buyer). Due to $Q_8$, we have $E_i[x^*] = v_i - x^*_{(2)} > 0 = E_i[x]$.

- **Case 7:** . $Q_1$, $\overline{Q}_3$ and $\overline{Q}_5$ for $x$, and $Q_2$, $\overline{Q}_4$ and $\overline{Q}_6$ for $x^*$ (the $i$-th buyer is among the $k$ highest buyers in both $x$ and $x^*$, he is not interested in paying $c$ and no other buyer is interested in paying $c$). Then, $E_i[x] = \frac{1}{k}\left(v_i - x_{(k+1)}\right)$ and $E_i[x^*] = \frac{1}{k}\left(v_i - x^*_{(k+1)}\right)$ (because $x_i > x_{(k+1)}$ ($Q_1$) and $v_i > x^*_{(k+1)}$, which implies that $x_{(k+1)} = x^*_{(k+1)}$)). Due to $Q_1$ and $Q_2$, we have

$$E_i[x^*] = \tfrac{1}{k}\left(v_i - x^*_{(k+1)}\right) = \tfrac{1}{k}(v_i - x_{(k+1)}) = E_i[x].$$

- **Case 8:** $Q_1$, $\overline{Q}_3$ and $Q_5$ for $x$, and $Q_2$, $\overline{Q}_4$ and $Q_6$ for $x^*$ (the $i$-th buyer is among the $k$ highest buyers in both $x$ and $x^*$; however, he is not interested in paying $c$, while someone else is interested in paying $c$). Then, $E_i[x] = E_i[x^*] = 0$.
- **Case 9:** $Q_1$, $\overline{Q}_3$ and $\overline{Q}_5$, then, $E_i[x] = \tfrac{1}{k}\left(v_i - x_{(k+1)}\right)$ (the $i$-th buyer is among the $k$ highest buyers in $x$; however, neither he nor any other buyer is interested in paying $c$). Also $Q_2$, $Q_4$ and $\overline{Q}_6$), then, $E_i[x^*] = v_i - c$ (the $i$-th buyer is among the $k$ highest buyers in $x^*$ and is interested in paying $c$, while no other buyer is interested in paying $c$). Due to $Q_1$ ($x_i > x_{(k+1)}$) and $Q_2$ ($v_i > x^*_{(k+1)}$), which implies $Q_4$ ($x_{(k+1)} = x^*_{(k+1)}$) and $v_i - c > \tfrac{1}{k}\left(v_i - x^*_{(k+1)}\right)$), we have $E_i[x^*] = v_i - c > \tfrac{1}{k}\left(v_i - x^*_{(k+1)}\right) = \tfrac{1}{k}(v_i - x_{(k+1)}) = E_i[x]$.
- **Case 10:** $Q_1$, $\overline{Q}_3$ and $Q_5$ (the $i$-th buyer is among the $k$ highest buyers in $x$, but he is not interested in paying $c$ and someone else is interested in paying $c$), then, $E_i[x] = 0$. Also $Q_2$, $Q_4$, $Q_6$ and $\overline{Q}_8$, then, $E_i[x^*] = 0$ (the $i$-th buyer is among the $k$ highest buyers in $x^*$, and he and others are interested in paying $c$, but he does not place the highest bid).
- **Case 11:** $Q_1$, $\overline{Q}_3$ and $Q_5$, then, $E_i[x] = 0$ (the $i$-th buyer is among the $k$ highest buyers in $x$, but he is not interested in paying $c$ and someone else is interested in paying $c$). Also $Q_2$, $Q_4$, $Q_6$ and $Q_8$, then, $E_i[x^*] = v_i - x^*_{(2)}$ (the $i$-th buyer is among the $k$ highest buyers in $x^*$, and both he and others are interested in paying $c$; furthermore, he places the highest bid). Due to $Q_8$, we have $E_i[x^*] = v_i - x^*_{(2)} > 0 = E_i[x]$.

**Table 4** Comparison between $E_i[x^*]$ and $E_i[x]$ when $x_i < v_i$

| Case | $Q_1$ | $Q_2$ | $Q_3$ | $Q_4$ | $Q_5$ | $Q_6$ | $Q_7$ | $Q_8$ | $E_i[x^*]$ | | $E_i[x]$ |
|---|---|---|---|---|---|---|---|---|---|---|---|
| 1 | No | No | No | No | ? | ? | No | No | 0 | = | 0 |
| 2 | No | Yes | No | No | ? | No | No | ? | $\tfrac{1}{k}\left(v_i - x^*_{(k+1)}\right)$ | > | 0 |
| 3 | No | Yes | No | No | ? | Yes | No | No | 0 | = | 0 |
| 4 | No | Yes | No | Yes | ? | No | No | Yes | $v_i - c$ | > | 0 |
| 5 | No | Yes | No | Yes | ? | Yes | No | No | 0 | = | 0 |
| 6 | No | Yes | No | Yes | ? | Yes | No | Yes | $v_i - x^*_{(2)}$ | > | 0 |
| 7 | Yes | Yes | No | No | No | No | ? | ? | $\tfrac{1}{k}\left(v_i - x^*_{(k+1)}\right)$ | = | $\tfrac{1}{k}(v_i - x_{(k+1)})$ |
| 8 | Yes | Yes | No | No | Yes | Yes | No | No | 0 | = | 0 |
| 9 | Yes | Yes | No | Yes | No | No | ? | Yes | $v_i - c$ | > | $\tfrac{1}{k}(v_i - x_{(k+1)})$ |
| 10 | Yes | Yes | No | Yes | Yes | Yes | No | No | 0 | = | 0 |
| 11 | Yes | Yes | No | Yes | Yes | Yes | No | Yes | $v_i - x^*_{(2)}$ | > | 0 |
| 12 | Yes | Yes | Yes | Yes | No | No | Yes | Yes | $v_i - c$ | = | $v_i - c$ |
| 13 | Yes | Yes | Yes | Yes | Yes | Yes | No | No | 0 | = | 0 |
| 14 | Yes | Yes | Yes | Yes | Yes | Yes | No | Yes | $v_i - x^*_{(2)}$ | > | 0 |
| 15 | Yes | Yes | Yes | Yes | Yes | Yes | Yes | Yes | $v_i - x^*_{(2)}$ | = | $v_i - x_{(2)}$ |

- **Case 12:** $Q_1$, $Q_3$ and $\overline{Q}_5$ for $x$, and $Q_2$, $Q_4$ and $\overline{Q}_6$ for $x^*$ (the $i$-th buyer is among the $k$ highest buyers in both $x$ and $x^*$, he is interested in paying $c$, and no other buyer is interested in paying $c$). Then, $E_i[x] = E_i[x^*] = v_i - c$.
- **Case 13:** $Q_1$, $Q_3$, $Q_5$ and $\overline{Q}_7$ for $x$, and $Q_2$, $Q_4$, $Q_6$ and $\overline{Q}_8$ for $x^*$ (the $i$-th buyer is among the $k$ highest buyers in $x$ and $x^*$, and in both vectors, he and others buyers are interested in paying $c$. However, the $i$-th buyer does not place the highest bid). Then, $E_i[x] = E_i[x^*] = 0$.
- **Case 14:** $Q_1$, $Q_3$, $Q_5$ and $\overline{Q}_7$, then, $E_i[x] = 0$ (the $i$-th buyer is among the $k$ highest buyers in $x$, and both, he and others are interested in paying $c$; however, he does not place the highest bid). Also $Q_2$, $Q_4$, $Q_6$ and $Q_8$, then, $E_i[x^*] = v_i - x^*_{(2)}$ (the $i$-th buyer is among the $k$ highest buyers in $x^*$ and both, he and others are interested in paying $c$, but he places the highest bid). Due to $Q_8$, we have $E_i[x^*] = v_i - x^*_{(2)} > 0 = E_i[x]$.
- **Case 15:** $Q_1$, $Q_3$, $Q_5$ and $Q_7$ for $x$, and $Q_2$, $Q_4$, $Q_6$ and $Q_8$ for $x^*$ (the $i$-th buyer is among the $k$ highest buyers in $x$ and $x^*$, and in both vectors, he and others players are interested in paying $c$, but he places the highest bid). Then, $E_i[x] = v_i - x_{(2)}$ and $E_i[x^*] = v_i - x^*_{(2)}$. Due to $Q_8$, which implies that $x_{(2)} = x^*_{(2)}$, we have $E_i[x^*] = v_i - x^*_{(2)} = v_i - x_{(2)} = E_i[x]$.

We found that for the 15 final cases the expected profit to the $i$-th buyer for $x^*$ is greater than or equal to that for $x$ and that the 15 final cases form a partition of $\Re^n_+$.

**Table 5** Comparison between $E_i[x^*]$ and $E_i[x]$ when $x_i > v_i$

| Case | $Q_1$ | $Q_2$ | $Q_3$ | $Q_4$ | $Q_5$ | $Q_6$ | $Q_7$ | $Q_8$ | $E_i[x^*]$ | | $E_i[x]$ |
|---|---|---|---|---|---|---|---|---|---|---|---|
| 1 | No | No | No | No | ? | ? | No | No | 0 | = | 0 |
| 2 | Yes | No | No | No | No | ? | ? | No | 0 | > | $\frac{1}{k}(v_i - x_{(k+1)})$ |
| 3 | Yes | No | No | No | Yes | ? | No | No | 0 | = | 0 |
| 4 | Yes | No | Yes | No | No | ? | Yes | No | 0 | > | $v_i - c$ |
| 5 | Yes | No | Yes | No | Yes | ? | No | No | 0 | = | 0 |
| 6 | Yes | No | Yes | No | Yes | ? | Yes | No | 0 | > | $v_i - x_{(2)}$ |
| 7 | Yes | Yes | No | No | No | No | ? | ? | $\frac{1}{k}(v_i - x^*_{(k+1)})$ | = | $\frac{1}{k}(v_i - x_{(k+1)})$ |
| 8 | Yes | Yes | No | No | Yes | Yes | No | No | 0 | = | 0 |
| 9 | Yes | Yes | Yes | No | No | No | Yes | ? | $\frac{1}{k}(v_i - x^*_{(k+1)})$ | > | $v_i - c$ |
| 10 | Yes | Yes | Yes | No | Yes | Yes | No | No | 0 | = | 0 |
| 11 | Yes | Yes | Yes | No | Yes | Yes | Yes | No | 0 | > | $v_i - x_{(2)}$ |
| 12 | Yes | Yes | Yes | Yes | No | No | Yes | Yes | $v_i - c$ | = | $v_i - c$ |
| 13 | Yes | Yes | Yes | Yes | Yes | Yes | No | No | 0 | = | 0 |
| 14 | Yes | Yes | Yes | Yes | Yes | Yes | Yes | No | 0 | > | $v_i - x_{(2)}$ |
| 15 | Yes | Yes | Yes | Yes | Yes | Yes | Yes | Yes | $v_i - x^*_{(2)}$ | = | $v_i - x_{(2)}$ |

Therefore, $E_i[x^*] \geq E_i[x] \quad \forall \ x \in \Re^n \ \text{such that} \ x_i < v_i,$ as desired. A summary is provided in Table 4. The case in which $x_i > v_i$ can be solved similarly; a summary for this case is provided in Table 5. □

## Appendix C: Algorithms

In this section we provide, in order of appearance, the pseudocode of the algorithms BU Myerson and the Draw auction method, in order to set light to the performance.

**Algorithm 1** BU Myerson *input*: $(a, b, \epsilon, p_1, p_2, x)$; *output*: $(\overline{p}, \overline{x})$

**Require:** $a, b, \epsilon, p_1, p_2, x$;
**Ensure:** $(x_{(1)}, \ldots, x_{(n)})$ = descending order of $x = (x_1, \ldots, x_n)$;
1: $\overline{p} = (\overline{p}_{(1)}, \ldots, \overline{p}_{(n)}) = (0, \ldots, 0)$;  $\overline{x} = (\overline{x}_{(1)}, \ldots, \overline{x}_{(n)}) = (0, \ldots, 0)$;
2: $A = \frac{p_1}{p_2}\frac{1-b}{a}$;  $B = -[\frac{p_1}{2a}(b - \frac{(1-b)}{p_2}(1 + p_1 + \epsilon)) + \frac{1}{2}]$;
3: $C = \frac{-a}{p_1}A^2 + \frac{1-b}{p_2}$;  $D = 2\frac{a}{p_1}AB$;  $E = \frac{-a}{p_1}B^2 + b - \frac{1-b}{p_2}(p_1 + \epsilon)$;
4: $y_1 = \frac{-D+\sqrt{D^2-4CE}}{2C}$; $y_2 = \frac{-D-\sqrt{D^2-4CE}}{2C}$; $z_1 = Ay_1 - B$; $z_2 = Ay_2 - B$;
5: $z_3 = (p_1 + \epsilon) - \sqrt{(p_1 + \epsilon)^2 - (p_1 + \epsilon) - \frac{p_1 b}{a}(p_1 + \epsilon - 1)}$;
6: $H_1(q) = \frac{a}{p_1}(q^2 - q)$; $H_3(q) = b(q-1) + \frac{1-b}{p_2}\left(q^2 - (1 + p_1 + \epsilon)q + (p_1 + \epsilon)\right)$;
7: **if** $\exists z_1 \in [0, p_1] \wedge \exists y_1 \in [p_1 + \epsilon, 1]$ **then**
8:   $z_0 := z_1$;  $y_0 := y_1$;
9: **else if** $\exists z_2 \in [0, p_1] \wedge \exists y_2 \in [p_1 + \epsilon, 1]$ **then**
10:  $z_0 := z_2$;  $y_0 := y_2$;
11: **else if** $\exists z_3 \in [0, p_1]$ **then**
12:  $z_0 := z_3$;  $y_0 := (p_1 + \epsilon)$;
13: **else**
14:  $z_0 := 0$;  $y_0 := (p_1 + \epsilon)$;
15: **end if**
16: $x_{min} = \frac{a}{2p_1}$;  $x_{\ell\ell} = \frac{az_o}{p_1}$;
17: $x_{cut} = F^{-1}(y_0)$;  $\beta_0 = \frac{H_3(y_0) - H_1(z_0)}{y_0 - z_0}$;
18: $\mathcal{D} = \{i \mid x_i \in [x_{\ell\ell}, x_{cut}]\}$;
19: **if** $\beta_0 < 0$ **then**
20:  **if** $x_{(1)} \geq x_{cut} \wedge x_{(2)} \in [0, x_{cut})$ **then**
21:   $\overline{p}_{(1)}(x) = 1$;  $\overline{x}_{(1)}(x) = x_{cut}$;
22:  **else if** $x_{(2)} \geq x_{cut}$ **then**
23:   $\overline{p}_{(1)}(x) = 1$;  $\overline{x}_{(1)}(x) = x_{(2)}$;
24:  **end if**
25: **else**
26:  **if** $x_{(1)} \notin [0, x_{min}) \wedge x_{(2)} \in [0, x_{min})$ **then**
27:   $\overline{p}_{(1)}(x) = 1$;  $\overline{x}_{(1)}(x) = x_{min}$;
28:  **else if** $x_{(1)}, x_{(2)} \in [x_{min}, x_{\ell\ell}]$ **then**
29:   $\overline{p}_{(1)}(x) = 1$;  $\overline{x}_{(1)}(x) = x_{(2)}$;
30:  **else if** $x_{(1)} \in (x_{\ell\ell}, x_{cut}] \wedge x_{(2)} \in [x_{min}, x_{\ell\ell}])$ **then**
31:   $\overline{p}_{(1)}(x) = 1$;  $\overline{x}_{(1)}(x) = x_{(2)}$;
32:  **else if** $x_{(1)}, x_{(2)} \in (x_{\ell\ell}, x_{cut}]$ **then**
33:   $\overline{p}_{(i)}(x) = \frac{1}{|\mathcal{D}|}, \forall i \in \mathcal{D}$;  $\overline{x}_{(i)}(x) = \frac{x_{\ell\ell}}{|\mathcal{D}|}, \forall i \in \mathcal{D}$;
34:  **else if** $x_{(1)} > x_{cut} \wedge x_{(2)} \in (x_{\ell\ell}, x_{cut}]$ **then**
35:   $\overline{p}_{(1)}(x) = 1$;  $\overline{x}_{(1)}(x) = x_{cut} - \frac{x_{cut} - x_{\ell\ell}}{|\mathcal{D}|+1}$;
36:  **else if** $x_{(1)} > x_{cut}$ **then**
37:   $\overline{p}_{(1)}(x) = 1$;  $\overline{x}_{(1)}(x) = x_{(2)}$;
38:  **end if**
39: **end if**

**Algorithm 2 Draw Auction** *input*: $(k, c, x)$; *output*: $(\overline{p}, \overline{x})$

**Require:** : $k, c, x$;
**Ensure:** $(x_{(1)}, \ldots, x_{(n)})$ = descending order of $x = (x_1, \ldots, x_n)$;
1: $\overline{p} = (\overline{p}_{(1)}, \ldots, \overline{p}_{(n)}) = (0, \ldots, 0)$; $\overline{x} = (\overline{x}_{(1)}, \ldots, \overline{x}_{(n)}) = (0, \ldots, 0)$;
2: **if** $x_{(2)} - c > \frac{1}{k}(x_{(2)} - x_{(k+1)})$ **then**
3:     $\overline{p}_{(1)}(x) = 1$; $\overline{x}_{(1)}(x) = x_{(2)}$;
4: **else if** $(x_{(2)} - c \leq \frac{1}{k}(x_{(2)} - x_{(k+1)})) \wedge (x_{(1)} - c > \frac{1}{k}(x_{(1)} - x_{(k+1)}))$ **then**
5:     $\overline{p}_{(1)}(x) = 1$; $\overline{x}_{(1)}(x) = c$;
6: **else if** $x_{(1)} - c \leq \frac{1}{k}(x_{(1)} - x_{(k+1)})$ **then**
7:     $D_k = \{i \mid x_i \in [x_{(1)}, \ldots, x_{(k)}]\}$;
8:     $\overline{p}_{(i)}(x) = \frac{1}{|D_k|}$, $\forall i \in D_k$;
9:     $\overline{x}_{(i)}(x) = \frac{x_{(k+1)}}{|D_k|}$, $\forall i \in D_k$;
10: **end if**

**Algorithm 3 Parameters estimation** *input*: $(a, b, p, n)$; *output*: $(k_{opt}, c_{opt}, saleValue_{opt})$

**Require:** : $a, b, p, n$;
**Ensure:** $k_{opt} = 0$; $c_{opt} = 0$; $saleValue_{opt} = 0$;
1: **for** $k = 1$ **to** $n - 1$ **do**
2:     **for** $c = 0$ **to** 1 **by** 0.01 **do**
3:         $average \leftarrow 0$;
4:         **for** $iter = 1$ **to** 1000000 **do**
5:             **for** $i = 1$ **to** $n$ **do**
6:                 $random \rightsquigarrow U(0, 1)$;
7:                 **if** $random < p$ **then**
8:                     $x_i \rightsquigarrow U(1, b)$;
9:                 **else**
10:                     $x_i \rightsquigarrow U(0, a)$;
11:                 **end if**
12:             **end for**
13:             $(\overline{p}, \overline{x}) \leftarrow$ Draw Auction$(k, c, x = (x_1, \ldots, x_n))$;
14:             $saleValue \leftarrow \sum_{\ell=1}^{n} \frac{\overline{x}_i}{p_i}$ (if $\frac{\overline{x}_i}{p_i} = \frac{0}{0}$, then $\frac{\overline{x}_i}{p_i} \leftarrow 0$);
15:             $average \leftarrow average + \frac{saleValue}{iter}$;
16:         **end for**
17:         **if** $average \geq saleValue_{opt}$ **then**
18:             $k_{opt} \leftarrow k$;
19:             $c_{opt} \leftarrow c$;
20:             $saleValue_{opt} \leftarrow saleValue$;
21:         **end if**
22:     **end for**
23: **end for**

**Author Contributions** All authors contributed to the study conception and design, as well as material preparation, data collection and analysis. The first draft of the manuscript was written by Javier Castro and Rosa Espínola, and all authors commented on previous versions of the manuscript. All authors read and approved the final manuscript.

**Funding** Open Access funding provided thanks to the CRUE-CSIC agreement with Springer Nature. This research has been partially supported by the Government of Spain, Grant Plan Nacional de I+D+i, PID2020-116884GB-I00, MTM2015-70550-P, PR108/20-28, PGC2018096509-B-I00 and TIN2015-66471-P, and the Complutense University of Madrid, PR108/20-28 and CT17/17-CT18/17.

## Declarations

**Competing Interests** The authors have no relevant financial or non-financial interests to disclose

Auctions. A New Method For Selling Objects... 1743## Authors and Affiliations

**Javier Castro[1,2] · Rosa Espínola[1,2] · Inmaculada Gutiérrez[1]  · Daniel Gómez[1,2]**

✉ Inmaculada Gutiérrez
  inmaguti@ucm.es

  Javier Castro
  jcastroc@estad.ucm.es

  Rosa Espínola
  rosaev@estad.ucm.es

  Daniel Gómez
  dagomez@estad.ucm.es

[1] Faculty of Statistical Studies, Complutense University, Av. Puerta de Hierro, sn., Madrid 28040, Spain

[2] Instituto de Evaluación Sanitaria, Complutense University, Ciudad Universitaria, Madrid 28040, Spain
Springer